\documentclass[twocolumn,floatfix,superscriptaddress,longbibliography,nofootinbib]{revtex4-2}
\RequirePackage[sort&compress]{natbib}
\usepackage[english]{babel}
\usepackage{color}
\usepackage{subcaption}
\usepackage{booktabs}
\usepackage{float}

\usepackage{amssymb,amsfonts,amsmath}
\usepackage{array}
\usepackage{bm}
\usepackage[pdftex]{graphicx}
\usepackage{xcolor}
\usepackage{comment}

\usepackage[colorlinks = true,
            linkcolor = blue,
            urlcolor  = blue,
            citecolor = blue,
            anchorcolor = blue]{hyperref}

\usepackage{dcolumn}
\usepackage{bm}
\usepackage{float}

\usepackage{xcolor}
\definecolor{RoyalBlue}{HTML}{4169e1}
\definecolor{ForestGreen}{HTML}{228b22}

\usepackage{setspace}
\usepackage{comment}

\usepackage{todonotes}

\newcommand{\ulysse}[1]{{\color{black} #1}}

\begin{document}


\title{Diffusive geodesics wandering in networks of rigid chains}


\author{Ulysse Marquis}
\email{ulyssepierre.marquis@unitn.it}
\affiliation{Fondazione Bruno Kessler, Via Sommarive 18, 38123 Povo (TN), Italy}
\affiliation{Department of Mathematics, University of Trento, Via Sommarive 14, 38123 Povo (TN), Italy}

\date{\today}

\begin{abstract}
    We introduce an ensemble of spatial networks built from the junctions of hindered-rotation chains, incorporating directional correlations between bonds, an aspect ignored in the standard network modeling paradigm. The emergent random networks support geodesics with a wandering exponent $\xi = 1/2$,  and a travel-time fluctuation exponent $\chi = 0$, consistent with the KPZ relation, yet violating the bound~$\chi\geq1/8$ predicted in the Poissonian framework. Transverse deviations follow the Kolmogorov distribution, indicating similarities between Brownian bridge excursions and geodesics in a random medium with correlated edges orientations. These results reveal a new universality class of Euclidean first-passage percolation, where local orientational memory reshapes transport properties and challenges existing bounds for random spatial networks.
\end{abstract}

\maketitle

\section{Introduction}

Most network models start from the assumption that connectivity emerges between point-like nodes, linked by pairwise relations or higher-order simplices~\cite{Albert_2002, Newman_2003, battiston2020networks}. Yet, in many real-world systems, the fundamental building blocks are not isolated nodes but chains or filaments—extended entities that branch, intersect, and form spatially embedded networks. Examples include biological and ecological systems and infrastructure, such as vascular networks~\cite{scianna2013}, river basins~\cite{dodds2000, rodriguez2017river}, fungal mycelia~\cite{Lee_2016}, transportation systems (metro lines, pipelines, railways)~\cite{Barthelemy_2011}, fracture and fault systems in materials~\cite{Andresen_2013}, fiber networks (paper, fabric)~\cite{Barabasi_Stanley_1995} and venation patterns in leaves~\cite{pelletier2000}, with sizes covering many scales. In these systems, the network's functionality is dependent on both its topology and its physical layout, characterized by a certain rigidity, i.e correlations between successive edges orientation. Such correlations have been overlooked in traditional network modeling--though recent work have started to emphasize the importance of the physical nature of networks~\cite{Dehmamy2018, Pósfai2024}.


By contrast, in soft-matter and polymer physics, the statistical mechanics of filamentary networks have long been developed to describe the formation and emergent properties of semiflexible polymers, gels, and rubbers~\cite{mackintosh1995elasticity, Rubinstein_2003, degennes1979scaling, flory1953principles, Broedersz_2014}. These frameworks successfully capture macroscopic transitions arising from microscopic processes (like cross-linking), leading to polymer network properties such as the elasticity of rubbers or the rigidity of gels. Ideal chain models from polymer theory, such as the worm-like chain~\cite{Rubinstein_2003}, provide a natural way to move beyond the node-based paradigm of spatial networks, introducing correlations between edge orientations in their spatial embedding.

The consequences of such directional correlations on global network properties remain underinvestigated. Of particular interest is their impact on the structure of shortest paths, which are fundamental to the study of complex networks, disordered systems physics \ulysse{and} other disciplines~\cite{bouttier2003}. In network science, shortest paths determine accessibility, centrality, and navigability~\cite{Barthelemy2022, Latora2001, Boguna2009}, while in statistical physics they underpin models of transport in random media such as first-passage percolation~\cite{auffinger201650yearspassagepercolation}. In disordered systems, the fluctuations of minimal paths reflect the underlying geometry of disorder: for example, interfaces such as domain walls in certain Ising models~\cite{HuseHenley1985_pinning} can be mapped to directed polymers in random media~\cite{derrida1990}, whose minimal energy configurations correspond to optimal paths~\cite{kardar1987}. Here, we extend this perspective to geometrically correlated chain networks, studying how directional correlations modify the \ulysse{shape} of shortest paths with Euclidean first-passage percolation~\cite{Howard1997}.

\ulysse{
In this work, we introduce a spatial network model built from the 
intersection of worm-like chains---filamentary objects whose 
successive bond orientations retain a local orientational memory. This 
construction departs from point-based spatial graphs and offers a way to control angular correlations of successive bonds. We first characterize the scaling of individual chains and analyze elementary geometric and topological properties of the resulting networks, 
including face statistics, global connectivity, and navigability. We then 
turn to the main result of the paper, the study of the shape of geodesics in the regime where chains are long and the network is densely interconnected.

Specifically, we examine the wandering of geodesics between 
points at Euclidean distance~$r$---whose transverse deviations 
scale as~$r^{\xi}$---as well as their deviations away from the mean 
length of the shortest path (also called travel-time fluctuations), which scale as~$r^{\chi}$. Simulations reveal that geodesics exhibit exponents $\xi = 1/2$ and $\chi = 0$ robustly across 
a range of small angular deviations, consistent with the predicted KPZ relation~$2\xi-1=\chi$, yet incompatible with lower bounds established in the Poissonian case. Moreover, transverse fluctuations collapse onto the Kolmogorov distribution characteristic of Brownian-bridge excursions, whereas travel-time fluctuations deviate from both Gaussian and Tracy--Widom forms. These results point to a new universality class of Euclidean first-passage percolation rooted in orientational memory and non-Poissonian spatial distribution of nodes.
}

\ulysse{The organization of the article is the following.} Section~\ref{sec:model} introduces the chain-network model, starting from the description of the construction of single chains (sec.\ref{subsec:chain}) and their scaling behavior (sec.\ref{subsec:scaling}), and proceeding to the analysis of macroscopic features of the network,such as the formation of a dense giant component (sec.~\ref{subsec:network}). In Section~\ref{sec:geodesics}, \ulysse{we introduce first-passage percolation and its Euclidean generalization, and discuss briefly the current state of the literature (sec.\ref{subsec:fpp}). Section~\ref{subsec:expe} discusses the experimental setup used to obtain the results. The measured exponents and fluctuations are described in sections~\ref{subsec:expo} and~\ref{subsec:fluctu}, before introducing a novel observable, the wiggliness, in section~\ref{subsec:wigg}.}





\section{Chain-network model}

\ulysse{
This section introduces the geometric and probabilistic ingredients of the model. We begin by describing how individual chains are generated as hindered–rotation polymers with parametrized angular persistence, and we characterize their scaling properties. We then explain how multiple chains, grown from randomly sampled seeds within a bounded domain, intersect to form a planar spatial network. Particular attention is given to identifying the regime in which this network becomes well-connected, as this is essential for the geodesic analysis of the next section. Additional network properties, which play a secondary role in the present study, are reported in the Appendix.
For the sake of clarity, Table~\ref{tab:symbols} summarizes the notations used throughout the paper.
}

\label{sec:model}

\begin{figure}[htp]
    \centering
    \begin{subfigure}[t]{0.49\linewidth}
        \centering
        \includegraphics[width=\linewidth]{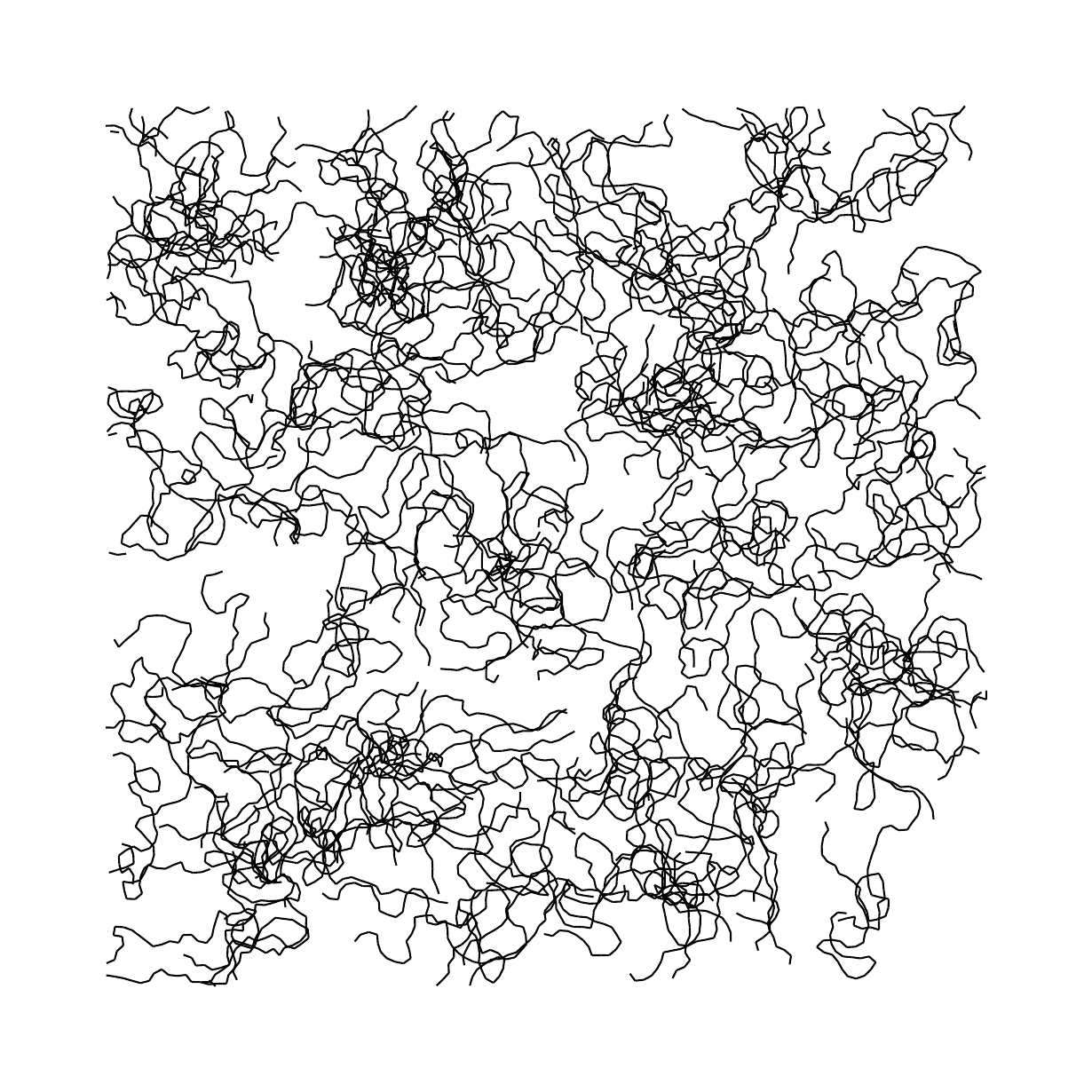}
        \caption{$L\approx \ulysse{\Delta}/2$, $\theta \gg 0$.}
    \end{subfigure}
    \hfill
    \begin{subfigure}[t]{0.49\linewidth}
        \centering
        \includegraphics[width=\linewidth]{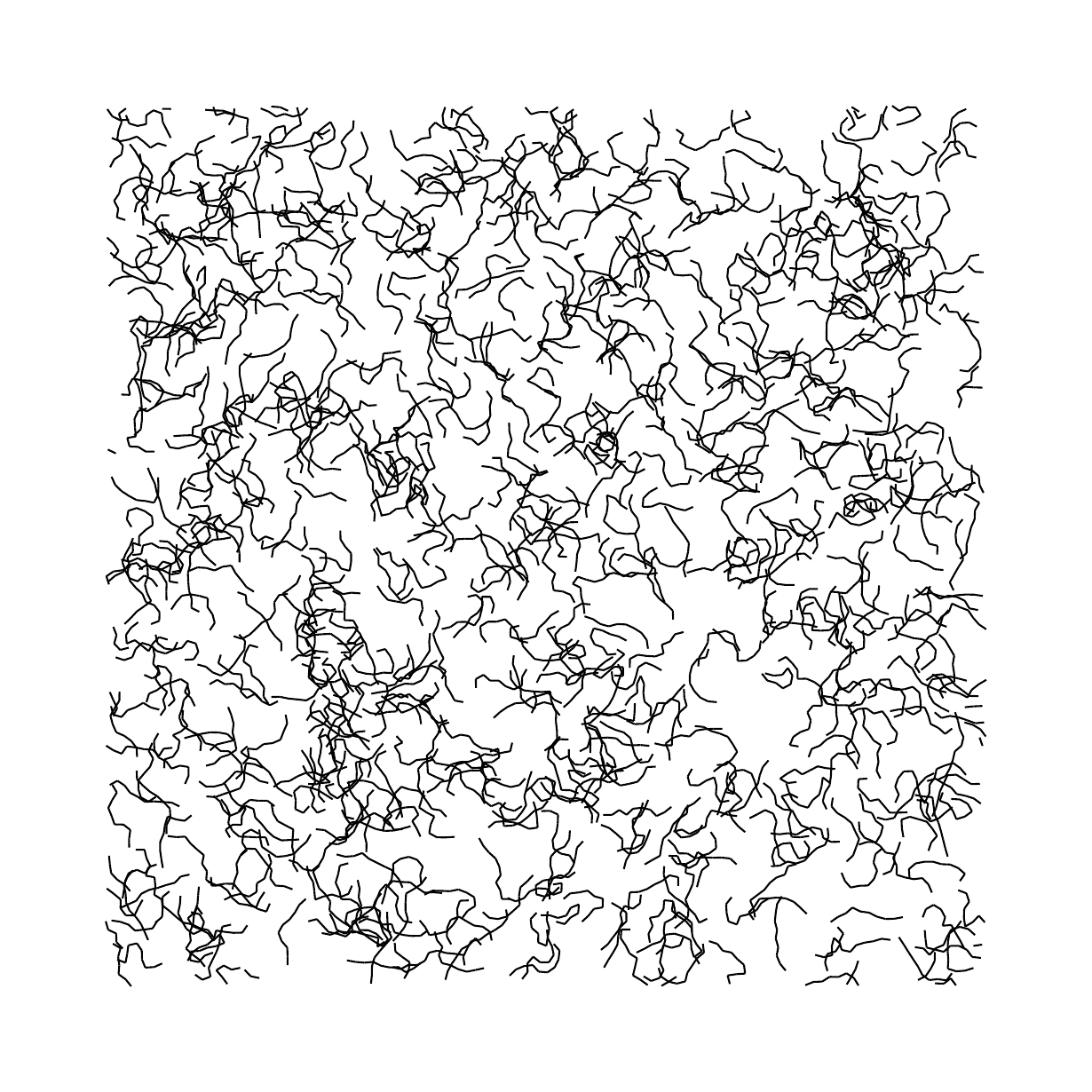}
        \caption{$L\approx \ulysse{\Delta}/10$,~$\theta \gg 0$.}
    \end{subfigure}

    \vspace{1em} 

    \begin{subfigure}[t]{0.49\linewidth}
        \centering
        \includegraphics[width=\linewidth]{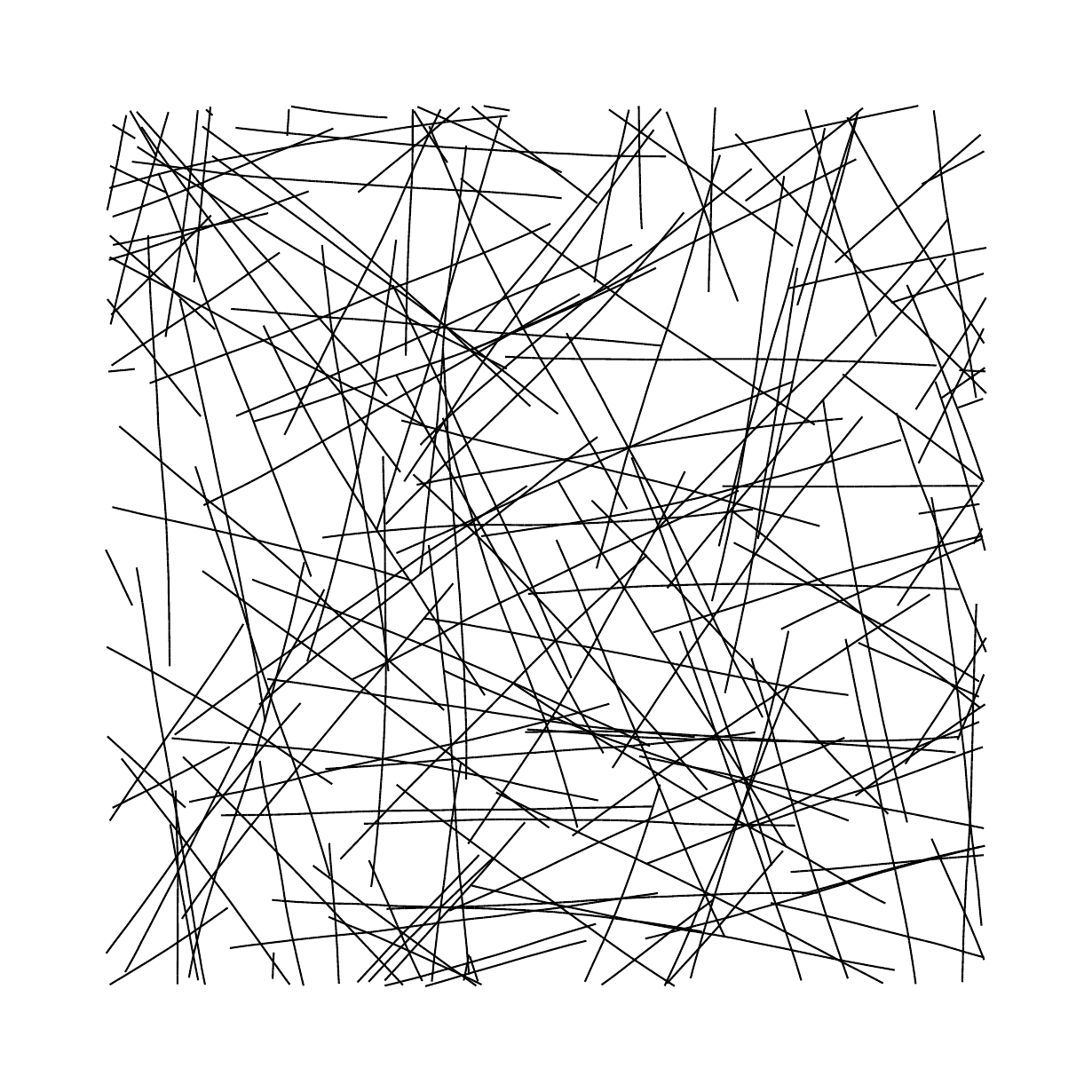}
        \caption{$L\approx \ulysse{\Delta}/2$,~$\theta=0$.}
        \label{fig:dense_straight_net}
    \end{subfigure}
    \hfill
    \begin{subfigure}[t]{0.49\linewidth}
        \centering
        \includegraphics[width=\linewidth]{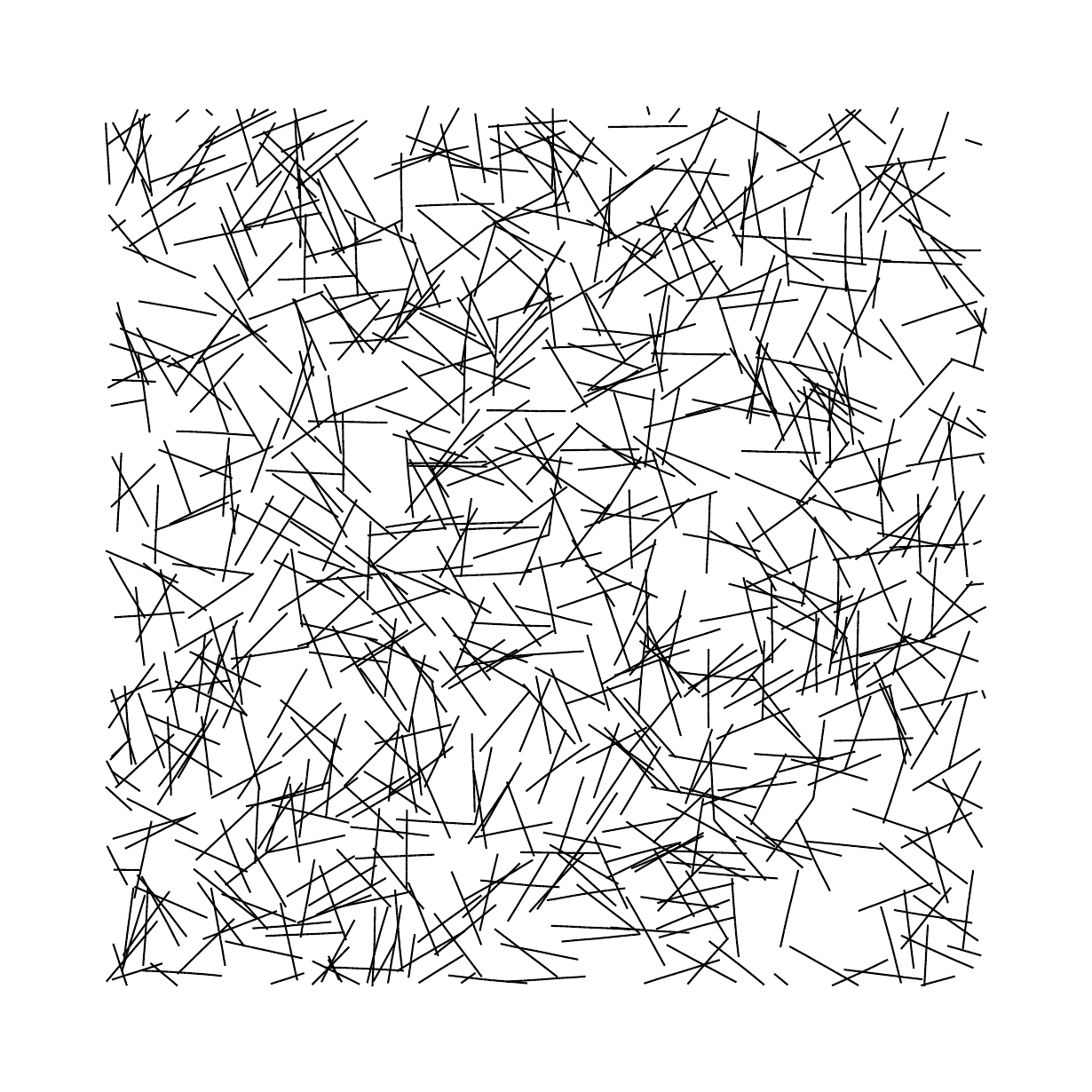}
        \caption{$L\approx \ulysse{\Delta}/10$,~$\theta=0$.}
    \end{subfigure}

    \caption{Realizations of networks with closed boundaries for various values of~$\theta$ and~$L$, at approximately equal density. The coupling between chain length and torsion is expected to affect macroscopic properties, such as the homogeneity of the chain length density. Configurations (a) and (c) show stronger heterogeneity in the domain coverage, while (b) and (d) cover it uniformly.}
    \label{fig:closed_boundaries}
\end{figure}

\ulysse{\subsection{Chain construction}\label{subsec:chain}}

The individual chains composing the network are constructed by the successive addition of~$L$ bonds of length~$\ell = 1$ attached to a random initial seed. The angular deviations between two consecutive bonds follow
\ulysse{
\begin{equation}
\phi_{i+1} = \phi_i + \eta_i ,
\end{equation}
}
where~$\langle \eta_i \rangle = 0$ and~$\langle \eta_i \eta_j \rangle = \theta^2 \delta_{i,j}$. Here~\ulysse{$\phi_i$} denotes the angle between bond~$i$ and the horizontal axis, in radians, while the initial orientation~\ulysse{$\phi_0$} is sampled \ulysse{uniformly at} random in\ulysse{~$[0,2\pi]$} \ulysse{and the~$\eta_i$ are sampled from a Gaussian of mean~$0$ and variance~$\theta^2$}. These chains are similar to correlated random walks (CRW) in continuous space~\cite{tojo96} and to ideal chains of the hindered-rotation model in polymer physics~\cite{Rubinstein_2003,doi1988}. 

\ulysse{\subsection{Single chain scaling}\label{subsec:scaling}}

\ulysse{
In polymer physics, chains are typically characterized by their radius of gyration
\begin{equation}
    R_g^{2} = \frac{1}{L+1}\sum_{i=1}^{L+1} |x_i - x_{\mathrm{cm}}|^{2} \,,
\end{equation}
with $x_i$ the monomer coordinates and $x_{\mathrm{cm}}$ their center of mass, and by their transverse length
\begin{equation}
    R_{\perp}^{2} = \frac{1}{L+1}\sum_{i=1}^{L+1} d_i^{2} \,,
\end{equation}
where $d_i$ denotes the orthogonal distance of monomer~$i$ to the chain’s end-to-end axis.
These two observables quantify the spatial extent and lateral fluctuations of the chain.

Typically, scaling exponents characterize how these quantities grow with chain length~$L$.
The Flory exponent~$\nu$ relates the radius of gyration to the polymer length through
\begin{equation}
    R_g \sim L^{\nu(\theta)} \,,
\end{equation}
and an analogous transverse exponent~$\nu_\perp$ is defined from the scaling of $R_\perp$.
By measuring $\nu$ and~$\nu_\perp$ as functions of the torsion parameter~$\theta$, we capture how chain geometry evolves as angular correlations evolve. Typically, angular correlations of chains are associated with a correlation length, called the persistence length~$\ell_P$~\cite{Rubinstein_2003}, which behaves as~$\ell_P\sim1/\theta^2$ when~$\theta\ll1$ and effectively quantifies the scale on which the orientation of bonds remain similar.

As~$\theta$ grows, the chains smoothly interpolate between two limiting regimes.
At very small~$\theta$, the behavior is rod-like (or worm-like), with exponents $\nu = 1$ and $\nu_{\perp} = 3/2$.
For large~$\theta$, the chains approach a correlated random walk, for which $\nu = \nu_\perp = 1/2$.
Our measured values (Figs.~\ref{fig:exponents_single} and~\ref{fig:nutranverse}) are consistent with classical results for semiflexible polymers and correlated random walks~\cite{Broedersz_2014, Rubinstein_2003, tojo96}.

The crossover between these regimes is smoother than the one reported in~\cite{barthelemy2025linesnetworks} for line-based networks with angular constraints: for example, at $\theta = 0.17\,\mathrm{ rad}$ we obtain a Flory exponent of about $0.7$, whereas a comparable angular constraint in~\cite{barthelemy2025linesnetworks} yields an exponent near $0.5$.
Finally, we emphasize that these exponents describe chains of large but finite length~$L$, which is precisely the regime relevant for the construction of the intersection networks studied here~\cite{barthelemy2025linesnetworks,tojo96}.
}

\begin{figure}[htp]
    \centering
    \includegraphics[width=1\linewidth]{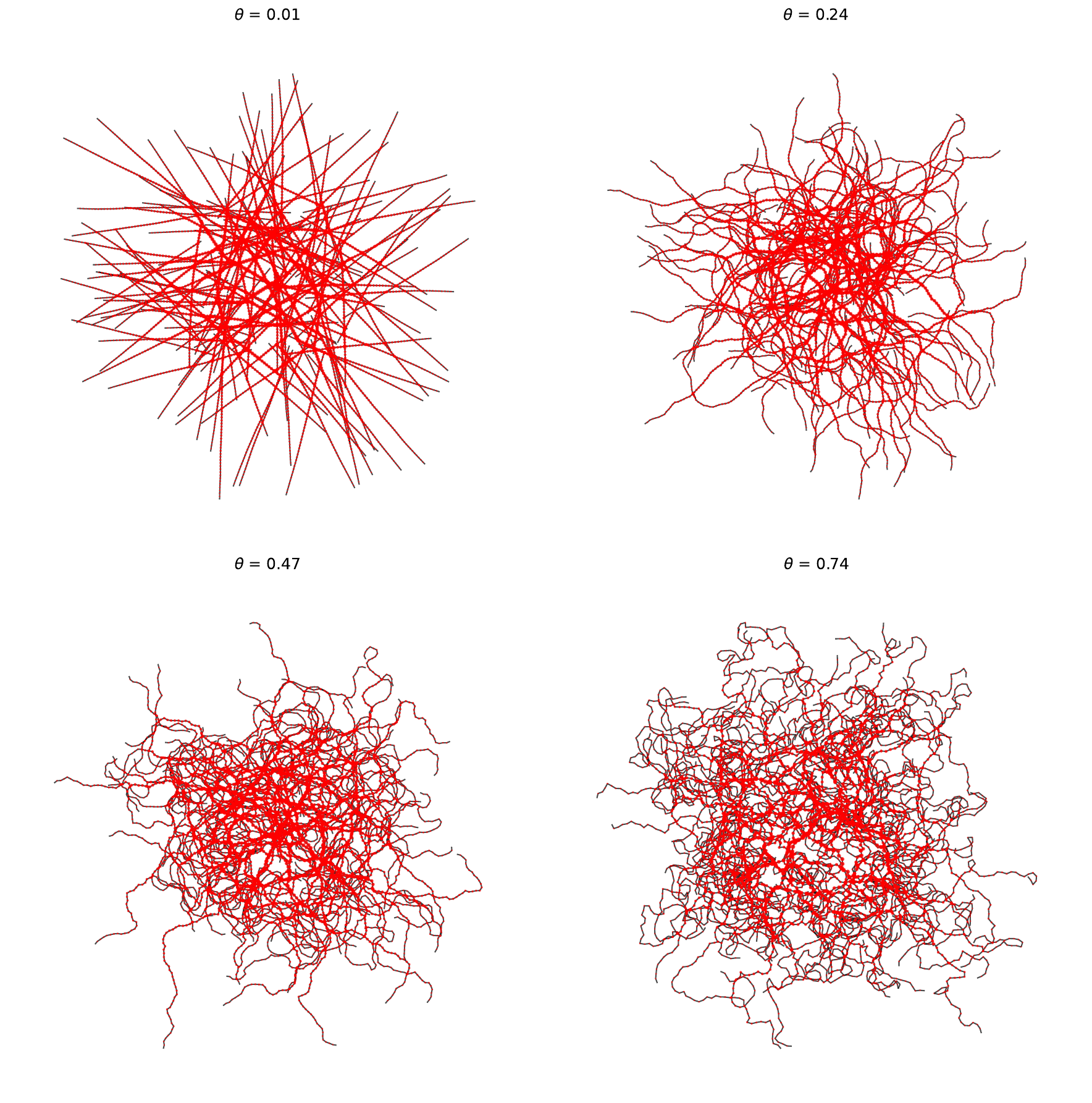}
    \caption{Network layout for~$L=50$,~$\ulysse{\Delta}=50$, open boundaries and various values of~$\theta$. Node size is proportional to their betweenness centrality \ulysse{(defined in Appendix~\ref{sec:navigability})}. As the torsion parameter increases, branches are less and less rigid, leading to the formation of peripheral loops.}
    \label{fig:fig:open_boundaries}
\end{figure}

\ulysse{\subsection{Network generation and properties} \label{subsec:network}}

\ulysse{
Chain networks are generated by sampling $M$ chains of length~$L$ whose bonds form the edges of a planar graph. Whenever two bonds intersect, a node is inserted at the intersection point, thus preserving planarity.
The initial seeds of the chains are sampled uniformly at random inside a bounded domain of linear extent~$\Delta$, with either closed or open boundary conditions. The present model generalizes earlier line-based systems such as the Mikado model
\cite{mackintosh1995elasticity,Broedersz_2014} and stick percolation 
\cite{jiantong2009}.

In the closed-boundaries configuration (Fig.~\ref{fig:closed_boundaries}), 
all chains are fully confined within the domain. 
More precisely, if a proposed segment would cross the boundary, 
the chain growth is halted at that point and no additional segments are added, 
resulting in a random distribution of  chain length with mean~$\mathrm{E}[L]$.  
In contrast, the open-boundaries configuration (Fig.~\ref{fig:fig:open_boundaries}) 
allows chains to extend freely beyond the domain. Instead, in the case of open boundaries, rod-like chains ($\theta \to 0$) tend to form a clear ``branches-and-core’’ structure, 
where long straight segments create a backbone with peripheral rectilinear branches.  
For random-walk-like chains (large $\theta$), the high curvature and frequent turns 
produce large external loops.  
Such structural differences are expected to influence functional properties of the system, 
including navigability, transport efficiency, and congestion.  

To investigate the geometry of shortest paths (Section~\ref{sec:geodesics}), 
we operate in a ``dense'' regime, in which the network is well-connected and spans spatially the domain.  
Because the continuum--percolation properties of this model~\cite{Meester_Roy_1996,sangare2009,Mertens_2012} are unknown---the chain length and torsion parameter~$\theta$ jointly affect connectivity---we determine the dense phase numerically.  
We measure the density of the giant component,
$$
    \mu = \frac{S}{\Delta^2},
$$
where $S$ is the total Euclidean length of the largest connected component, 
as a function of the chain-length density
$$
    \rho = \frac{M\,\mathrm{E}[L]}{\Delta^2}.
$$
For both rigid rods ($\theta=0$) and highly flexible chains ($\theta=0.74$), 
we find a critical density $\rho_c \simeq 2$ above which $\mu > 1$, 
provided chains are sufficiently long (Fig.~\ref{fig:perco_combined}). The geometric properties of the faces in this dense phase are analyzed in 
Section~\ref{sec:faces}, while navigability and transport aspects are discussed in 
Section~\ref{sec:navigability}.
}

\ulysse{
In the remainder of the paper, in order to study the geodesics of the well-connected, spanning phase of the ensemble, we focus on the regime~$\rho>\rho_c$ with chain lengths of order the system size. }

\begin{table}[H]
\centering
\footnotesize
\begin{tabular}{l|l}
\toprule
Notation & Definition \\
\midrule

$\phi_i$ & Orientation of bond $i$ (angle) \\
$\theta$ & Chain stiffness parameter  \\
$L$ & Length of chains\\
$\Delta$ & Linear size of the square domain \\
$R_{\perp}$ & Chain transverse length \\
$R_g$ & Chain radius of gyration \\
$\nu$ & Radius of gyration scaling exponent \\
$\omega$ & Fluctuation of radius of gyration scaling exponent \\
$\nu_\perp$ & Transverse length scaling exponent \\
$S$ & Length of the giant connected component \\
$\mu$ & Giant component density $S / \Delta^{2}$ \\
$\rho$ & Chain length density~$M\mathrm{E}[L]/\Delta^2$\\
$\rho_c$ & Density above which~$\mu>1$ \\
$r$ & Euclidean distance between nodes, $r = |x-y|$ \\
$T(x,y)$ & Shortest-path travel-time between nodes $x$ and $y$ \\
$D(x,y)$ & Transverse deviation of the geodesic between $x$ and $y$ \\
$V(T)$ & Variance of the travel time $T$ \\
$\mathrm{E}[D]$ & Expected transverse deviation  \\
$\delta$ & Normalized fluctuations of~$D$ \\
$t$ & Normalized fluctuations of~$T$ \\
$\chi$ & $V(T)$ scaling exponent,~$V(T) \sim r^{2\chi}$\\
$\xi$ & $\mathrm{E}[D]$ scaling exponent,~$\mathrm{E}[D] \sim r^{\xi}$ \\
$\gamma$ & Geodesic or path \\
$\ell(\gamma)$ & Euclidean length of~$\gamma$ \\
$W_\gamma$ & Wiggliness of~$\gamma$ \\
$\langle . \rangle$ & Ensemble average \\
$\ell_P$ & Chain persistence length (see~\cite{Rubinstein_2003} for a precise definition) \\

\bottomrule
\end{tabular}
\caption{List of notations used throughout the paper. The dependency in the distance~$r$ of~$V(T)$ and~$\mathrm{E}[D]$ is dropped.}
\label{tab:symbols}
\end{table}

\section{Geodesics}
\label{sec:geodesics}

\ulysse{
The goal of this section is to study the geodesics of the network in the regime where chains are long and the network is well-connected and spanning space. We begin by introducing the mathematical framework and reviewing relevant literature. Next, we present the details of the experiment conducted on the network model, followed by the results and their interpretation. Finally, we propose the \textit{wiggliness}, an observable inspired by polymer physics that could provide insight into the geometry of geodesics. 
}

\subsection{First-Passage Percolation}
\label{subsec:fpp}

\ulysse{
First-Passage Percolation (FPP)~\cite{Hammersley1965} provides a minimal model for random transport in disordered media and offers a natural framework to study the geometry of geodesics. In FPP, non-negative random weights are assigned to the edges of a lattice~$\mathbb{Z}^d$, and the geodesic between two points $x$ and $y$ is defined as the path $\gamma$ minimizing the total travel time
\begin{equation}
    T(x,y) = \min_{\gamma: x \to y} \ell(\gamma),
\end{equation}
where $\ell(\gamma)$ is the sum of the weights along the edges of $\gamma$. While the edge weights are independent, $T(x,y)$ is a minimum over correlated sums of these weights, and thus exhibits non-trivial fluctuations.  

Our main interest lies in the \emph{shape} of geodesics and the statistical properties of these fluctuations. On large scales, the mean travel time grows linearly with the Euclidean distance, $T(x,y) \sim |x-y|$, giving the first-order description of the random metric ball, which converges to a limiting, compact shape as formalized by the shape theorem~\cite{cox1981,auffinger201650yearspassagepercolation}. Beyond the average behavior, the transverse deviations of the geodesics encode the geometry of the paths. We define the maximal orthogonal deviation $D(x,y)$ as the largest perpendicular distance between the geodesic and the straight line connecting $x$ and $y$. In the language of disordered systems, one can then characterize the geodesic ensemble using the expectation $\mathrm{E}[D]$ and the variance of travel times $\mathrm{V}(T) = \mathrm{E}[(T-\mathrm{E}[T])^2]$, which capture the typical fluctuations around the mean shape. Note here that, unless needed, we drop the explicit~$x,y$ dependence of~$T$ and~$D$ in order to lighten the notations.

When these quantities obey power-law scaling with distance,
\begin{equation}
    \mathrm{V}(T) \sim |x-y|^{2\chi}, \qquad \mathrm{E}[D] \sim |x-y|^{\xi},
\end{equation}
the exponents $\chi$ (travel-time fluctuation) and $\xi$ (wandering) are conjectured to satisfy the Kardar–Parisi–Zhang (KPZ) relation~\cite{HuseHenley1985_pinning,kardar1987,chatterjee2012universalrelationscalingexponents},
\begin{equation} \label{eq:kpz}
    \chi = 2\xi - 1.
\end{equation}
In two dimensions, the expected values $\xi = 2/3$ and $\chi = 1/3$ place FPP in the KPZ universality class, with fluctuations conjecturally converging to Tracy–Widom statistics~\cite{Calabrese_2010,takeuchi2010,Halpin_Healy_2015}.  

Euclidean FPP (EFPP)~\cite{Howard1997} generalizes this framework to random spatial networks. In EFPP, points are drawn from a Poisson point process and pairs of points are connected by edges weighted by a power of their Euclidean distance. The finite-$\alpha$ regime exhibits rich scaling behavior, with rigorous bounds $\frac{1}{8} \leq \chi \leq \frac{1}{2}$ and $\xi \leq \frac{3}{4}$~\cite{Howard_2000,howard2000geodesicsspanningtreeseuclidean}. EFPP has been further extended to various spatial graphs including supercritical percolation clusters~\cite{Brereton_2014}, Delaunay triangulations~\cite{PIMENTEL_2011}, and proximity graphs~\cite{hirsch2015,Kartun_Giles_2019}, where the edge weights effectively reduce to the Euclidean distances. Remarkably, \cite{Kartun_Giles_2019} found that two families of scaling exponents consistently appear across these networks, $(\chi,\xi) = (1/5,3/5)$ for proximity graphs and $(2/5,7/10)$ for excluded-region graphs~\cite{Barthelemy2022}, both satisfying the KPZ relation (\ref{eq:kpz}) but exhibiting Gaussian, rather than Tracy–Widom, fluctuations.  

In this work, we adopt this EFPP perspective to focus on the shape of geodesics in networks formed by long chains.
}

\begin{figure}[htp]
    \centering
    \includegraphics[width=1\linewidth]{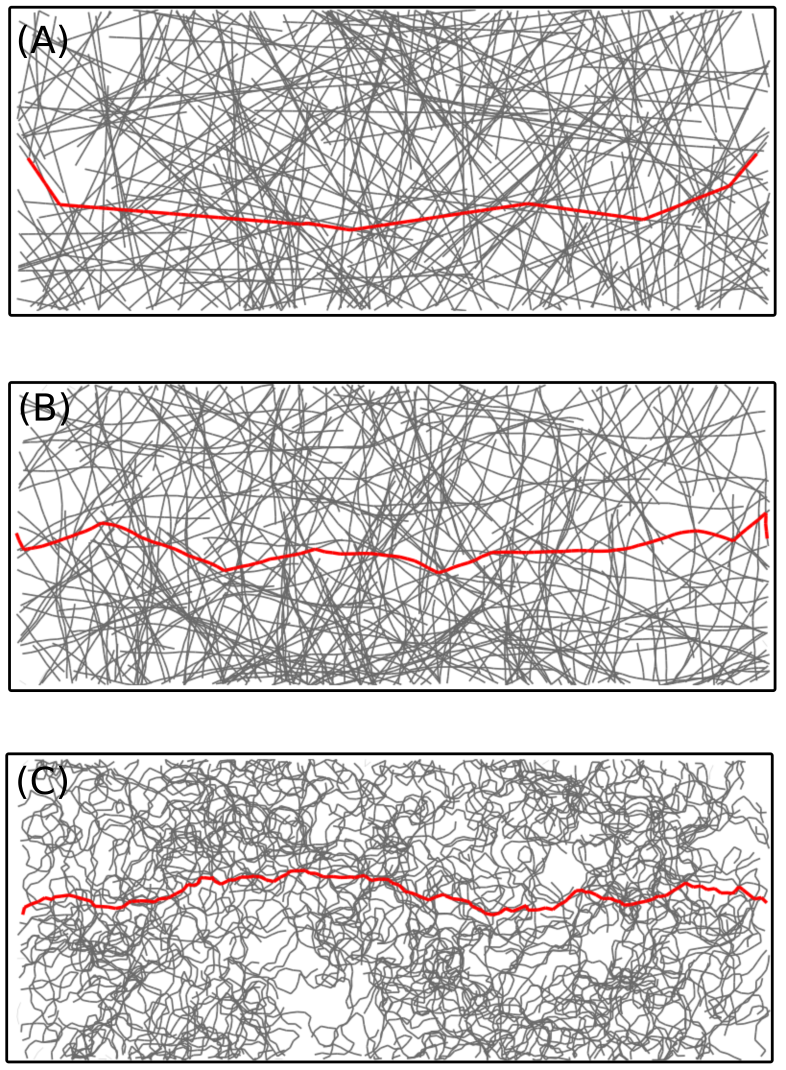}
    \caption{Realizations of the model for~$\theta=0$ (A),~$\theta=0.09$ (B) and~$\theta=0.74$ (C), for line density~$\rho=3$. The red lines represent geodesics from points close to~~$(0,h/2)$ and~$(W,h/2)$. Notice in  (A) and (B) the long `straight' subsequences of the geodesics, corresponding to navigation on a single chain. Since the persistence length (see Appendix~\ref{sec:scaling}) of these chains is very high--it grows in~$\theta^{-2}$--navigation along a single chain occurs almost rectilinearly. }
    \label{fig:geodesics}
\end{figure}

\begin{figure*}[htp]
    \centering
    \includegraphics[width=1\linewidth]{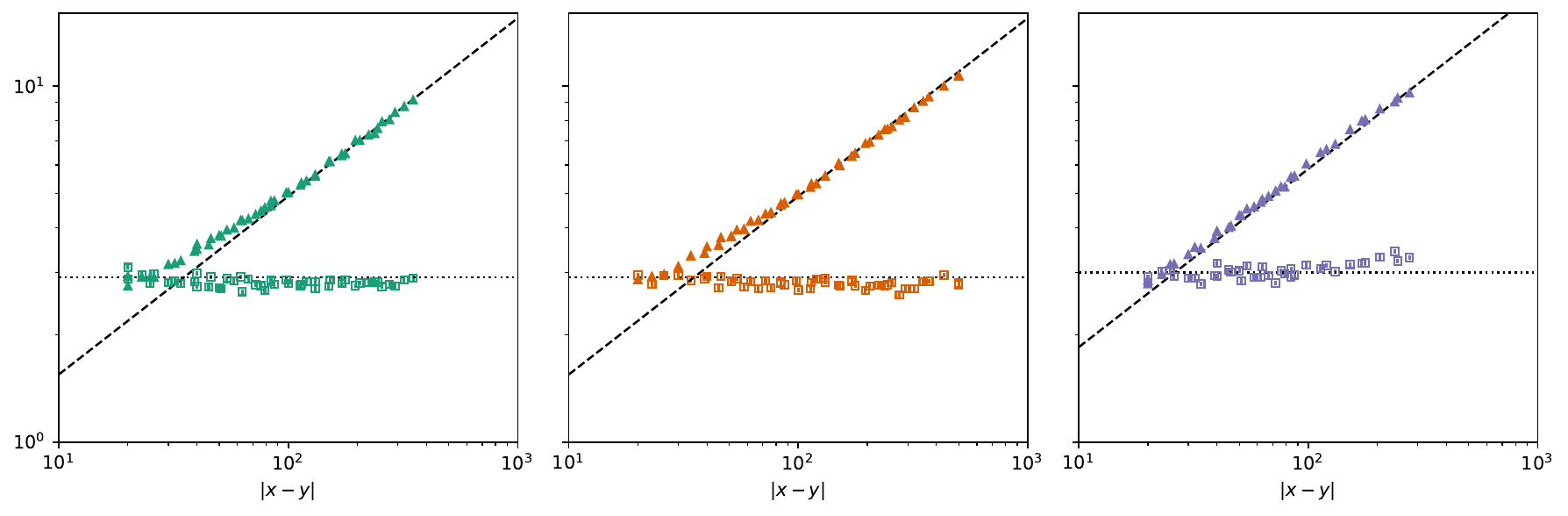}
    \caption{Scaling of the average transversal deviation~$\mathrm{E}(D)$ (triangles) and of~$V(T)^{1/2}$ (squares) in function of the Euclidean distance for~$\theta=0$ (left),~$\theta=0.09$ (center) and~$\theta=0.74$ (right). Each point represents an average over~$2000$ realizations, the dashed line is a power-law of exponent~$1/2$, and the pointed line is a constant, plotted as a guide for the eye. }
    \label{fig:tranverse_scaling}
\end{figure*}

\subsection{Experimental setup}
\label{subsec:expe}

\ulysse{
To measure the travel time $T(x,y)$ and the transverse deviation $D(x,y)$, we reproduce the experimental setup of~\cite{Kartun_Giles_2019}. Specifically, within the closed domain 
$$
[-W/2-b,\, W/2+b] \times [-h/2-\epsilon,\, h/2+\epsilon],
$$ 
where $b$ and $\epsilon$ are small buffer widths, we sample a density $\rho =3 > \rho_c$ of chains. Two additional chains of length~$L$ are introduced with seeds located at $(-W/2,0)$ and $(W/2,0)$. The height $h$ is chosen sufficiently large so as not to artificially constrain the transverse deviation $D$, i.e., always more than six standard deviations away from $\mathrm{E}[D]$. The domain width $W$ is varied between $20$ and $1000$ to explore the scaling behavior of $V(T)$ and $\mathrm{E}[D]$, although computational limitations restrict the maximal width to roughly $W\approx 300$–$600$. The chain length $L$ is taken of the system linear size to emulate effectively infinite chains crossing a finite domain, avoiding small-size artifacts (e.g., short chains appearing as sticks when zoomed out).  

For each configuration, the shortest path between the two points~$(-W/2,0)$ and~$(W/2,0)$ is computed, using the Dijkstra algorithm, with each edge weighted by their Euclidean length, yielding sequences of observables $\{D_i(r)\}$ and $\{T_i(r)\}$. Given the continuous nature of the shortest path length, there is only one minimum for each pair of node with probability~$1$. For each $W$, we sample 2000 independent configurations and retain only one geodesic per graph to prevent correlations that could decrease the precision of the estimates~\cite{Kartun_Giles_2019}.  


Here, we present results for three representative cases: perfectly aligned rods ($\theta=0$), slightly perturbed rods ($\theta=0.09$), and a smooth correlated random walk ($\theta\approx0.74$). The choice $\theta=0.09$ is motivated by its role as a small perturbation of the $\theta=0$ case, which in many physical systems can lead to qualitative changes in behavior. Indeed, this small torsion already modifies the geometric properties of the chains, producing exponents that are intermediate between the two limiting cases. For the rods, we have $\nu=1$ and $\nu_\perp=3/2$, while for $\theta=0.74$, $\nu\approx\nu_\perp\approx0.5$, and for $\theta=0.09$ the exponents fall between these extremes. The scaling behavior for intermediate values, $0.1 \leq \theta \leq 0.7$, is presented in Appendix~\ref{sec:intermed}. From the measured values of $T_i(r)$ and $D_i(r)$, we compute the empirical variance $V(T(r))$ and mean $\mathrm{E}[D(r)]$, and extract scaling exponents via linear fits in log–log scale, which also allows for estimation of confidence intervals on the exponents~\cite{Montgomery2012}. For $\theta = 0$ and $0.09$, the fits are restricted to $r>10^2$. A similar approach is adopted for the case~$0.1 \leq \theta \leq 0.7$ in the Appendix~\ref{sec:intermed}.

The main challenge of this experiment lies in the computational cost: even with an efficient implementation, the time required to sample graphs grows rapidly with $W$, compounded by the sensitivity of the measurements~\cite{Kartun_Giles_2019}. The experimental setup is illustrated in Fig.~\ref{fig:geodesics}. For rigid chains, the geodesics (red lines) often consist of subsequences following individual chains, leading to minimal angular deviations.}

\subsection{Results : exponents}
\label{subsec:expo}

Fig.~\ref{fig:tranverse_scaling} presents the measured average transversal deviations~$\mathrm{E}[D]$ and the travel-time standard deviation~$\mathrm{V}(T)^{1/2}$ in function of the Euclidean distance~$\ulysse{r=|x-y|}$. For rod-like systems ($\theta = 0$ and $\theta = 0.09$), the scaling behavior emerges only after an initial transition regime at~$r \leq 10^2$, whereas for the smoother correlated random walk ($\theta = 0.74$) the scaling relation seems to also hold at~$r \leq 10^2$. Further experiments \ulysse{(shown in Appendix~\ref{sec:intermed})} reveal that this crossover \ulysse{disappears smoothly as~$\theta$ increases}. For all the considered angles, the scaling exponent is close to~$1/2$, see Table~\ref{tab:scaling_exponents} for the measured values. The variance of the travel time~$\mathrm{V}(T)$, shows no power-law increase with distance, i.e~$\chi=0$, consistent with the KPZ relation~$2 \xi-1=\chi$. This exponent value means that the path show no considerable increase of their `roughness' (as can be seen on the Fig.~\ref{fig:geodesics}, where paths are locally straight lines) as the distance between points increases. It suggests that~$\mathrm{V}(T)$ is either independent of~$r$ or grows sub–power-law (for instance, logarithmically), with measurable effects only at much larger scales. \ulysse{These results also hold in the regime~$0.1 \leq \theta \leq 0.7$, as shown in the Appendix~\ref{sec:intermed}.}


The measured values of the fluctuation exponent~$\chi$ breaks the lower bound~$\chi=1/8$ predicted by Howard~\cite{Howard_2000} for complete graphs on Poissonian point processes and raises questions about generalizations of EFPP results on non-Poissonian graphs and about the existence of universality classes outside of the Poissonian paradigm~\cite{howard2000geodesicsspanningtreeseuclidean,Howard1997,Howard_2000}. 
\begin{table}[h!]
\centering
\begin{tabular}{l c c}
\hline
\textbf{Configuration} & $\xi$ & $\chi$  \\
\hline
$\theta=0$ & $0.49\pm0.03$ &  $0.00 \pm 0.03$\\
$\theta=0.09$  & $0.49\pm0.01 $   & $0.00\pm0.02$  \\
$\theta=0.74$   & $0.50\pm0.01$   & $0.02\pm0.02$ \\
\hline
\end{tabular}
\caption{\ulysse{Measured scaling exponents $\xi$ for geodesic transverse deviations and $\chi$ for travel time fluctuations, \ulysse{obtained from ordinary least-squares fits in log–log scale}, with $95\%$ confidence intervals, estimated using standard inferential statistics methods~\cite{Montgomery2012}. The reference values $\chi=0$ and $\xi=1/2$ lie within the error bars in all cases.}
}
\label{tab:scaling_exponents}
\end{table}

\begin{figure}[htp]
    \centering
    \includegraphics[width=0.9\linewidth]{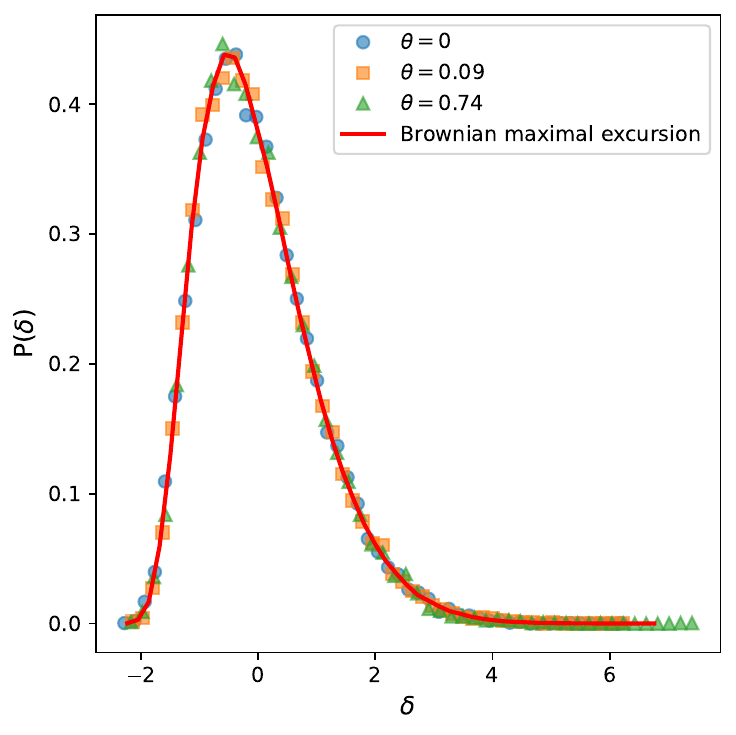}
    \caption{\ulysse{Distribution of normalized} fluctuations~$\delta$ of the transverse width~$D$, for the three torsion parameters considered. The red line corresponds to the rescaled distribution of the maximum of Brownian bridge excursions, defined in~\ref{eq:bb}.}
    \label{fig:fluctuations_transverse}
\end{figure}

\subsection{Results : fluctuations}
\label{subsec:fluctu}

Higher-order moments and full fluctuation distributions are typically relevant in statistical physics of disordered systems~\cite{Halpin_Healy_2015}. Fig.~\ref{fig:fluctuations_transverse} displays the normalized fluctuations of the transversal deviations $\delta$, \ulysse{defined as follows: the empirical average of the observed $\{D_i(r)\}$ is subtracted from each value, and the resulting deviations are divided by their empirical standard deviation, thereby allowing to compare fluctuations at different~$r$.} The resulting distributions collapse onto a single curve and are well-described by the Kolmogorov distribution $K$, corresponding to the maximum of a Brownian bridge $B$ on $[0,1]$,  
\begin{equation} \label{eq:bb}
    K = \sup_{0 < t < 1} |B(t)| \,.
\end{equation}
\ulysse{A Brownian bridge on~$[0,\tau]$ is a standard Brownian motion~\cite{Rogers_Williams_2000} conditioned to start and end at zero over the interval $[0,\tau]$, i.e., $B(0)=B(\tau)=0$.}

This collapse onto the Kolmogorov distribution is surprising, as the exponent~$1/2$ also corresponds to the transversal deviation exponent if the geodesics where exactly Brownian bridges. Furthermore, measurements of the standard deviation of~$D$ yield exponents of approximately~$0.40$ (see Fig.~\ref{fig:std_transverse}). While this deviates slightly from the theoretical value of~$1/2$ expected for \ulysse{the variance of the} maximum transversal deviation of Brownian bridges~\cite{marsaglia2003}, the difference is not large enough to rule out this possibility, especially considering the limited scale of the experiment. Moreover, the Kolmogorov fluctuations differ from the fluctuations observed in~\cite{Kartun_Giles_2019} on many spatial networks, including planar ones.

\begin{figure}[htp]
    \centering
    \includegraphics[width=0.9\linewidth]{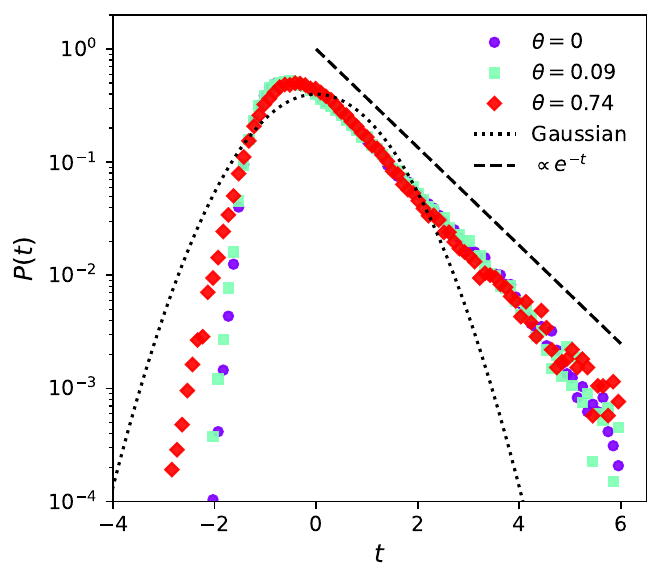}
    \caption{\ulysse{Normalized} travel time fluctuations~$t$ (colored points) ; pointed line : standard Gaussian distribution ; dashed line : exponential distribution. \ulysse{For all shown values of~$\theta$, the distribution is positively skewed.} The three distributions collapse on the right tail, while the configuration at high~$\theta$ shows a a noticeable difference with the rods-like chains in the left-tail.}
    \label{fig:fluctuations_traveltime}
\end{figure}

 In Fig.~\ref{fig:fluctuations_traveltime} the \ulysse{distribution of normalized travel time} fluctuations~$t$ \ulysse{(computed analogously to~$\delta$)} are displayed. The fluctuations do not collapse on a single master curve, independent of~$\theta$, as it is the case for~$\delta$. Moreover, they are not Gaussian, as their right tail decreases like~$\exp(-t)$ and their left-tail is sub-Gaussian. They also are not Tracy-Widom (TW), as shows their skewness \ulysse{which} is stable in the range~$[1.2,1.4]$, far above the TW skewness values, for any ensemble. \ulysse{This interval is consistently above but close to the skewness of the Gumbel distribution,~$\gamma \approx1.14$, suggesting some similarity with extreme-value statistics.}  
 
 These results hint towards the existence a different class of first-passage travel time fluctuations when the orientations of consecutive edges are correlated. Extensive simulations are needed to assess the dependence of the scaling exponents \ulysse{and of the distribution of fluctuations} with the torsion angle~$\theta$. In particular, the existence of a critical point, beyond which the correlations are weak enough and the geodesics fall back to the usual universality classes, is in question.
 
\subsection{Wiggliness}
\label{subsec:wigg}

In the light of polymer physics, we measured the \textit{wiggliness} of the geodesics, defined as
\begin{equation}
    W_\gamma = \frac{1}{\ell(\gamma)} \sum_{i=0}^{|\gamma|-1} (\phi_{i+1} - \phi_i)^2 \,,
\end{equation}
where~$\ell(\gamma)$ is the length of the geodesic~$\gamma$, and~$\phi_{i+1} - \phi_i$ are the successive bond angular differences along~$\gamma$. This quantity can be interpreted as a discrete version of the average bending energy density~\cite{Rubinstein_2003, Broedersz_2014} for the geodesics.

The measurements of the average wiggliness on the networks shown in Fig.~\ref{fig:geodesics} reveal (a) little dependence on the Euclidean distance between nodes, and (b) a non-trivial, yet expected, dependence on the torsion parameter~$\theta$ (see Tab.~\ref{tab:wiggliness}). By decomposing~$\gamma$ into sequences of paths that navigate along sub-chains, one can approximate~$\langle W_\gamma \rangle \approx \ell(\gamma)^{-1} \left( s \theta^2 + r \Xi^2 \right)$, where~$s$ quantifies the number of hops that follow a single chain, characterized by an average torsion~$\theta$, while~$r$ quantifies the number of jumps between distinct chains, associated with an average angular deviation~$\Xi$. There is, a priori, no direct information about~$\Xi$, although—since~$\gamma$ is a geodesic—a tendency for consecutive chains to align is naturally expected.

In particular, this observable may serve to estimate a \textit{persistence length} of the geodesics, $\ell_P(\gamma) \sim 1/W_\gamma$~\cite{Rubinstein_2003}. The Delaunay graph yields $\langle W_\gamma \rangle \approx 1.87 > 1$, indicating no effective directional memory of the geodesics, while all chain-networks give $\langle W_\gamma \rangle < 1$. We therefore propose this observable as a means to distinguish broad classes of geodesics on planar graphs.

\begin{table}[h!]
\centering
\begin{tabular}{cc}
\hline
\textbf{Configuration} & \textbf{$\langle W_\gamma \rangle$} \\
\hline
$\theta = 0.00$ & $0.11\pm0.17$ \\
$\theta = 0.09$ & $0.12\pm0.12$ \\
$\theta = 0.74$ & $0.87\pm0.24$ \\
Delaunay & $1.87\pm0.36$ \\
\hline
\end{tabular}
\caption{Values of the average wiggliness $\langle W_\gamma \rangle$, \ulysse{$\pm$ the standard deviation of the distribution} for different torsion parameter~$\theta$ and the Delaunay triangulation graph, at equivalent density.}
\label{tab:wiggliness}
\end{table}

\section{Conclusion}

We have proposed a spatial network model based on the junction of polymer-like chains, introducing correlations between bond orientations, ignored in the traditional paradigm. It is only parametrized by the length of the chains, the torsion parameter~$\theta$ and the number of chains, making it a useful baseline for chain-based networks. Our analysis reveals a new universality class of Euclidean first-passage percolation (EFPP) on correlated chain networks, characterized by~$\chi= 0$ and~$\xi = 1/2$ -- a regime that respects the KPZ relation while it violates the standard bounds derived for Poissonian graphs. Remarkably, the transverse fluctuations collapse onto the Kolmogorov distribution, while travel-time statistics deviate from both Gaussian and Tracy–Widom forms.

Future work should map out the phase diagram of this model through large-scale simulations. In this study, we have focused on relatively small torsion parameters and limited spatial scales, constrained by computational cost. A central open question is to identify the boundary between correlated and uncorrelated regimes, where orientational memory is lost and the system reverts to the familiar Poissonian universality classes. This crossover is likely governed by the \textit{persistence length}~$\ell_P \sim 1/\theta^2$, which sets the characteristic scale over which bond orientations remain correlated.

Overall, these results demonstrate that directional correlations--an intrinsic feature of chain-based networks--act as a distinct source of geometric disorder, reshaping the scaling laws of transport and shortest-path geometry. By borrowing concepts from polymer physics, the chain-network model provides a minimal framework to explore how local orientational memory reorganizes the network global geometry. In this light, chain-networks emerge as a new frontier for random geometries induced by spatial networks, challenging existing universality classes and motivating a broader theory of transport in disordered media.

\subsection*{Acknowledgements}
UM is thankful to Marc Barthelemy for the discussions about chains crossings.


\subsection*{Data Availability}
The data supporting the findings were sampled using the code available at~\cite{Marquis2025_chain_network_sampler}.





\markboth{}{}

\bibliographystyle{apsrev4-2}

\bibliography{biblio_ulysse}

\appendix

\renewcommand{\thefigure}{A\arabic{figure}}
\setcounter{figure}{0}

\renewcommand{\thesection}{A\arabic{section}} 
\setcounter{section}{0}   

\clearpage
\newpage 

\section{Scaling of single chains} \label{sec:scaling}

Firstly, we discuss the building block of our networks. Polymer physics has proposed a number of models of ideal chains~\cite{Rubinstein_2003}. The freely-rotating chain does not account for the variation of potentials and all torsion angles have equal probability. More realistic models are the worm-like chain, similar to the freely-jointed chain but with very small torsion angles~$\theta$, with the persistence length~$\ell_P \sim 1/\theta^2$ ($\ell_P$ is defined as the correlation length for the orientation of chains, see~\cite{Rubinstein_2003}), and the hindered rotation model, for which torsion angles~$\phi$ are determined by a Boltzmann factor~\cite{Rubinstein_2003, doi1988}
\begin{equation} 
    \langle \cos \phi \rangle =\frac{\int \cos \phi \, \exp(-U/kT) \, d\phi}{\int \, \exp(-U/kT) \, d\phi} \,.
\end{equation}
Similarly, the correlated random walk (CRW) in continuous space~\cite{tojo96} considers the construction of polymers like the freely-rotating chains, where the consecutive torsion angles are sampled uniformly in~$[-\theta/2,\theta/2]$.~\cite{tojo96} showed that chains of length~$L$ are characterized by a short-length regime
\begin{equation} \label{eq:flory}
    R_g \sim L^{\nu(\theta)} \,,
\end{equation}
where~$1/2 \leq \nu \leq 1$ is akin to the Flory exponent in polymer physics~\cite{flory1953principles}. Equation~\ref{eq:flory} holds until an upper length scale~$L_c \sim \theta^{-h}$, with~$h \approx 1.88$. At~$L \gg L_c$,
\begin{equation}
    R_g^2 = f(\theta) L \,,
\end{equation}
the random walk regime is retrieved.

In this work, we propose the following model, an adaptation of the CRW in continuous space inspired from the hindered rotation model. An initial bond of length~$a=1$ with random angle~$\phi_0$ is chosen. Then, the chain is grown for~$L-1$ segments of of length~$a$, with consecutive angles following the relation
\begin{equation}
    \ulysse{\phi_{i+1} = \phi_i + \eta_i \,,}
\end{equation}
with~$\langle \eta_i \rangle = 0$ and~$\langle \eta_i \eta_j \rangle = \theta^2 \delta_{i,j}$. Specifically, the~$\eta_i$ are distributed as a Gaussian.  We characterize the chains of the model by computing their Flory exponents~\cite{doi1988,flory1953principles,degennes1979scaling,Rubinstein_2003}, their fluctuations
\begin{equation}
    \langle R_g^2 - \langle R_g \rangle^2\rangle^{1/2} \sim L^{\omega(\theta)} \,,
\end{equation}
and their transversal width
\begin{equation}
    R_\perp \sim L^{\nu_\perp} \,.
\end{equation}

\begin{figure*}[htp]
    \centering
    \includegraphics[width=1\linewidth]{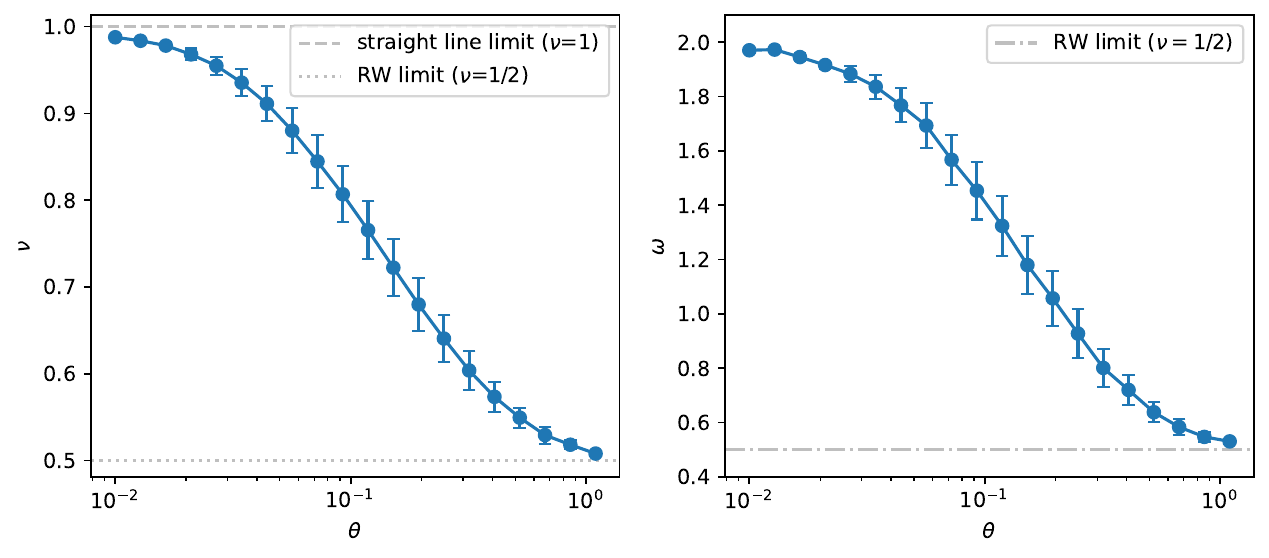}
    \caption{Flory exponent~$\nu$ and fluctuation exponent~$\omega$ in function of~$\theta$. Exponents crossover from rods-like (or worm-chain like values) at ~$\theta\sim0$ to the random walk regime~$\nu=\nu_\perp=1/2$. Values of exponents were measured of~$10^3$ realizations of chains ranging from~$L=20$ to \ulysse{~$L=10^4$} (far from an eventual~$L_c$~\cite{tojo96}). Error bars represent the~$95\%$ confidence intervals. Grey lines represent special limits, the rods ($\theta=0$) and the RW.}
    \label{fig:exponents_single}
\end{figure*}

\begin{figure}[htp]
    \centering
    \includegraphics[width=1\linewidth]{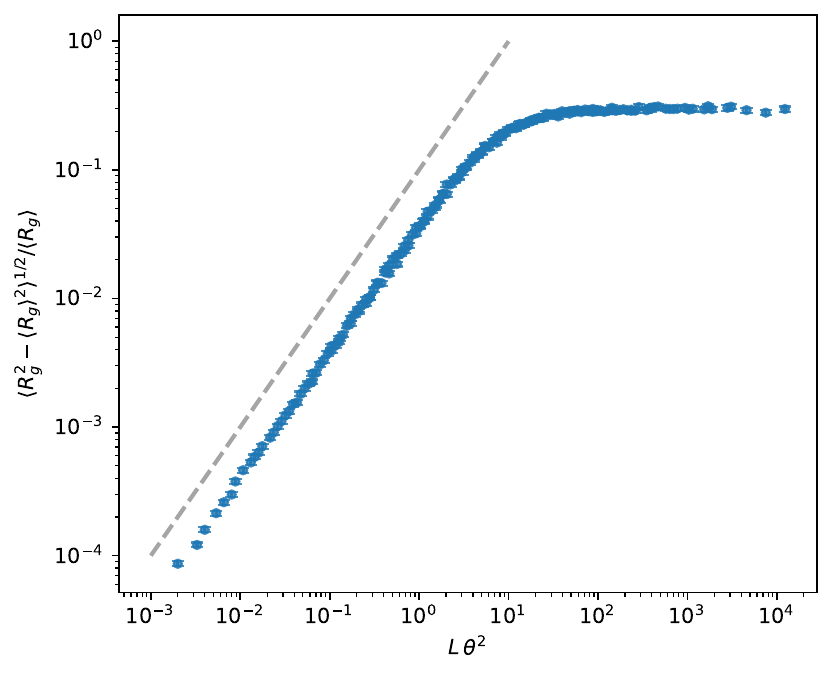}
    \caption{Relative fluctuations in function of~$L\theta^2$, the rescaled length. When the chain length is smaller than its persistence length~$L\theta^2 \ll 1$, the relative fluctuations grow as a linear function of the rescaled length chain, before saturating around~$L \theta^2=1$ (if~$L$ grows, at~$L_c\sim1/\theta^2$, and if~$\theta$ grows, at~$\theta_c \sim L^{-1/2}$. Grey dashed line :~$L\theta^2$. Each point (and confidence interval) comes from the realizations described in Fig.~\ref{fig:exponents_single}.}
    \label{fig:relativefluct}
\end{figure}

We measured the Flory and fluctuation exponents of the proposed chains, while staying in the regime~$L < L_c$ mentioned above. Their values are shown in Fig.~\ref{fig:exponents_single}. For the Flory exponent, we observe a crossover between the rod regime~$R_g \sim L$, i.e~$\nu=1$ and the random walk regime~$\nu=1/2$. The random walk fluctuations~$\omega=1/2$ are also retrieved when~$\theta\gg 0$, while at small disorder parameter~$\theta\approx0$, we find the surprisingly high exponent~$\omega\approx2$. This can be explained by looking at the relative fluctuations in function of the ratio~$L/\ell_P \sim L\theta^2$, see Fig~\ref{fig:relativefluct}. We observe the collapse of the relative fluctuations on a master curve~$f$,
\begin{equation}
    \frac{\langle R_g^2 - \langle R_g \rangle^2 \rangle^{1/2}}{\langle R_g \rangle} = f(L \theta^2) \,,
\end{equation}
characterized by~$f(u \ll 1) \sim u$ and~$f(u \gg 1)\sim1$. When~$\theta\to0$, as~$\ell_P$ grows like~$\theta^{-2}$, an exponent~$\omega=1+\nu$ appears as an artifact of the regime~$L \ll \ell_P$.

\begin{figure}[htp]
    \centering
    \includegraphics[width=1\linewidth]{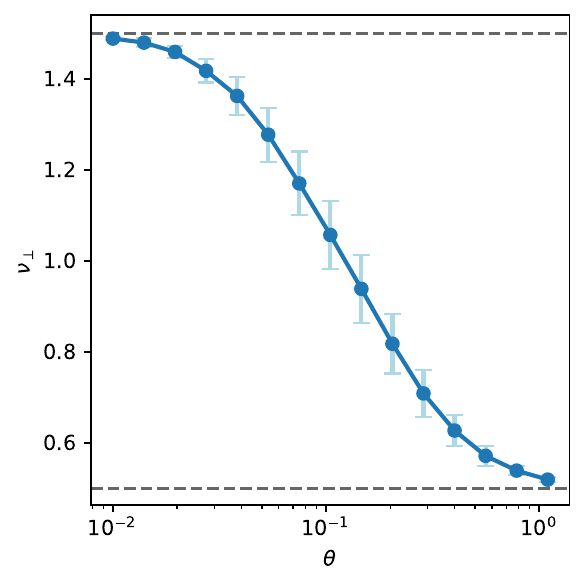}
    \caption{Transversal exponent~$\nu_\perp$ in function of~$\theta$. The worm-like chain regime is retrieved as~$\theta\to0$, with an exponent~$3/2$~\cite{Rubinstein_2003}, while the random walk regime~$\nu_\perp=1/2$ appears at large~$\theta$. In between, there is a crossover between these exponents. For instance, at~$\theta\approx10^{-1}$,~$\nu_\perp\approx1$. The experimental setup is the same as in Fig.~\ref{fig:exponents_single}.}
    \label{fig:nutranverse}
\end{figure}

In the small-angle limit ($\theta \to 0$), the polymer exhibits the worm-like chain scaling~\cite{Broedersz_2014} with $\nu_\perp = 3/2$, whereas for large angular fluctuations ($\theta \gg 0$) it crosses over to the diffusive regime of the random-walk with $\nu_\perp = 1/2$, as shown in Fig.~\ref{fig:nutranverse}.


\clearpage

\clearpage

\section{Giant component}

\begin{figure*}[b]
    \centering
    \begin{minipage}{0.48\linewidth}
        \centering
        \includegraphics[width=\linewidth]{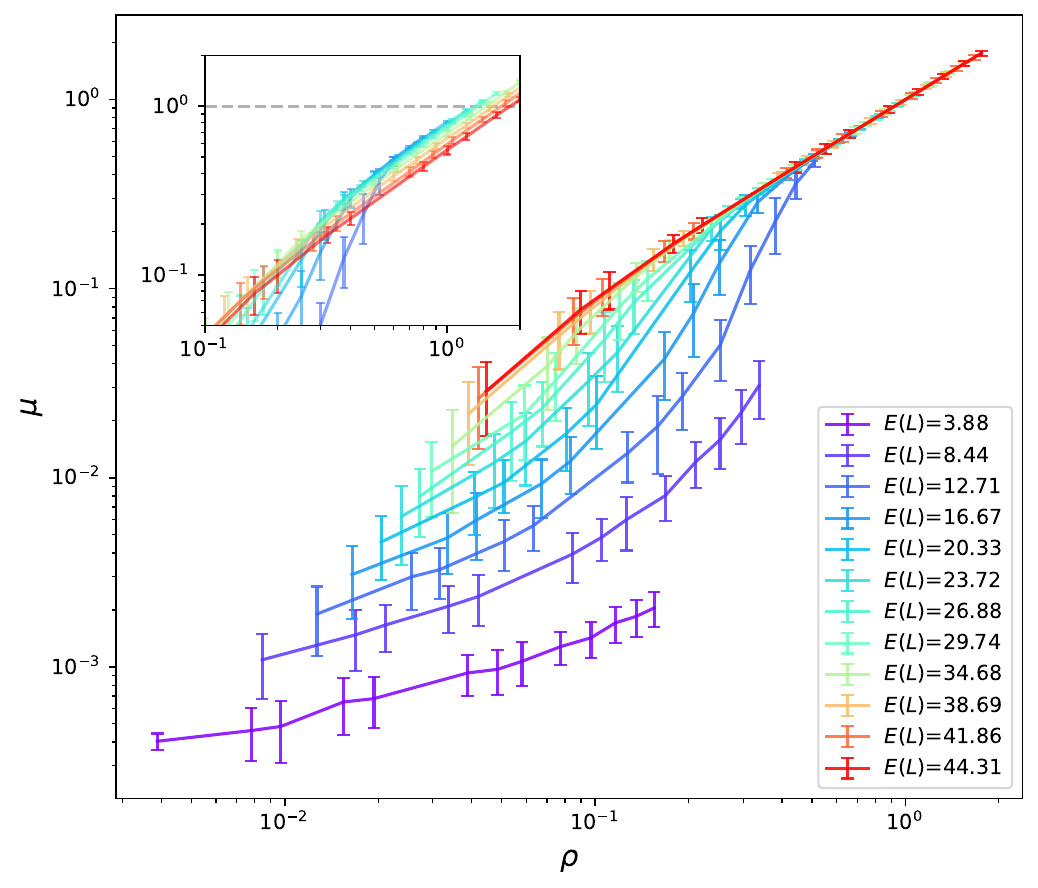}
        \label{fig:perco_rods}
    \end{minipage}
    \hfill
    \begin{minipage}{0.48\linewidth}
        \centering
        \includegraphics[width=\linewidth]{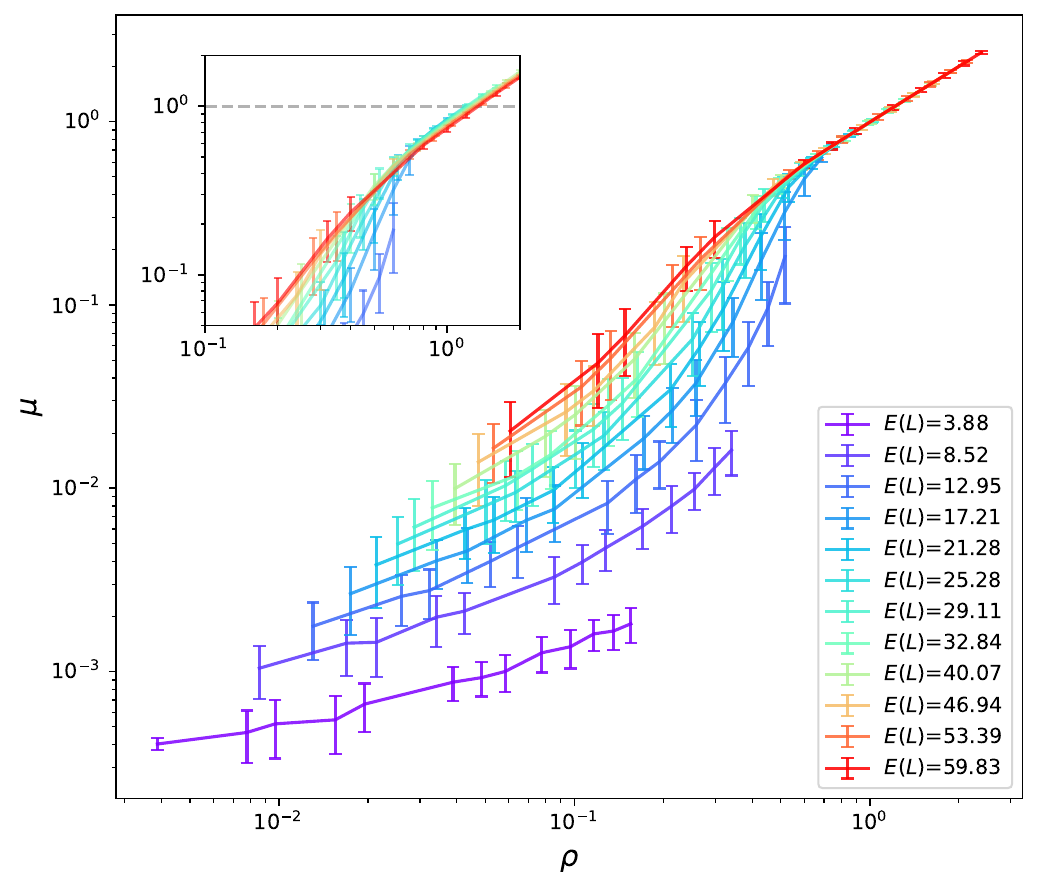}
        \label{fig:perco_rw}
    \end{minipage}
    \caption{Percolation analysis for~$\theta=0$ (left) and~$\theta=0.74$ (right), in the case of fixed~$\ulysse{\Delta}=100$. All points are averaged over~$100$ realizations, with the error bars indicating the~$95\%$ confidence intervals. Inset : zoom on the high~$\rho$ part, dashed line :~$\mu=1$. }
    \label{fig:perco_combined}
\end{figure*}

While the goal of the paper is not the percolation analysis of the chain-network model, it is necessary to understand roughly which density of chains is required for obtaining a giant component which also spans the domain. We analyze the cases of rods~$\theta=0$ and smooth CRW ($\theta=0.74$), and vary~$M$ and~$L$. Results are shown in Fig.~\ref{fig:perco_combined}. We notice firstly that the curves do not collapse on one master curve, as would happen if~$\mu$ was only a function of~$\rho$, but there is a dependency in~$\mathrm{E}(L)$. At high~$\rho$ values, all curves organize around one behavior, where~$\mu$ grows linearly with~$\rho$, indicating the appearance of a giant component absorbing a large fraction of lines. We see in the insets that between~$\rho=1$ and~$\rho=2$, depending on~$L$ and~$\theta$, the giant component density overpasses~$\mu=1$.

\clearpage

\section{Faces statistics} \label{sec:faces}

\begin{figure}[H]
    \centering
    \includegraphics[width=1\linewidth]{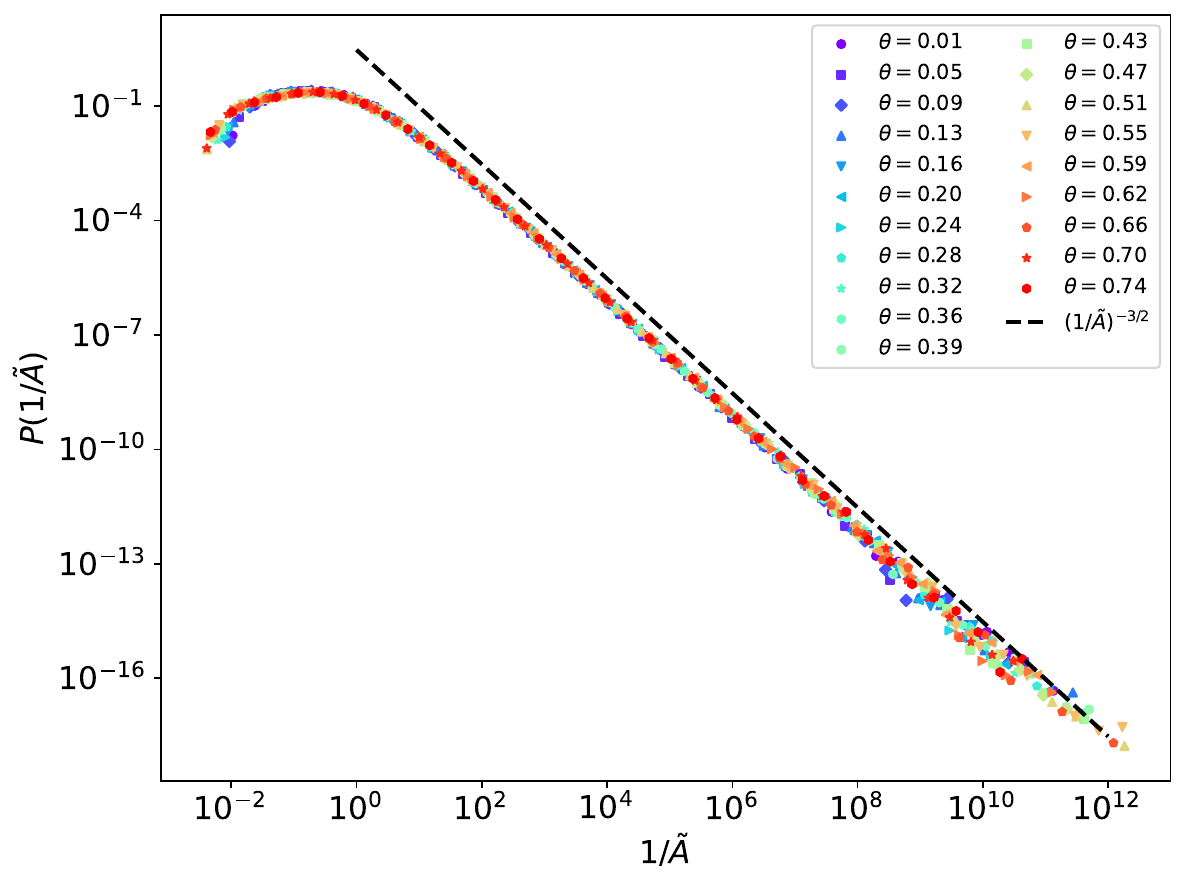}
    \caption{Distribution of inverse rescaled areas. Dashed line : power-law of exponent~$-3/2$. Areas statistics were obtained for~$100$ realizations for~$20$ values of~$\theta$ (values indicated in the legend), in the open boundaries setup -- which should have the statistics for small faces than the closed boundaries model.  }
    \label{fig:area_universal}
\end{figure}

\begin{figure}[H]
    \centering
    \includegraphics[width=1\linewidth]{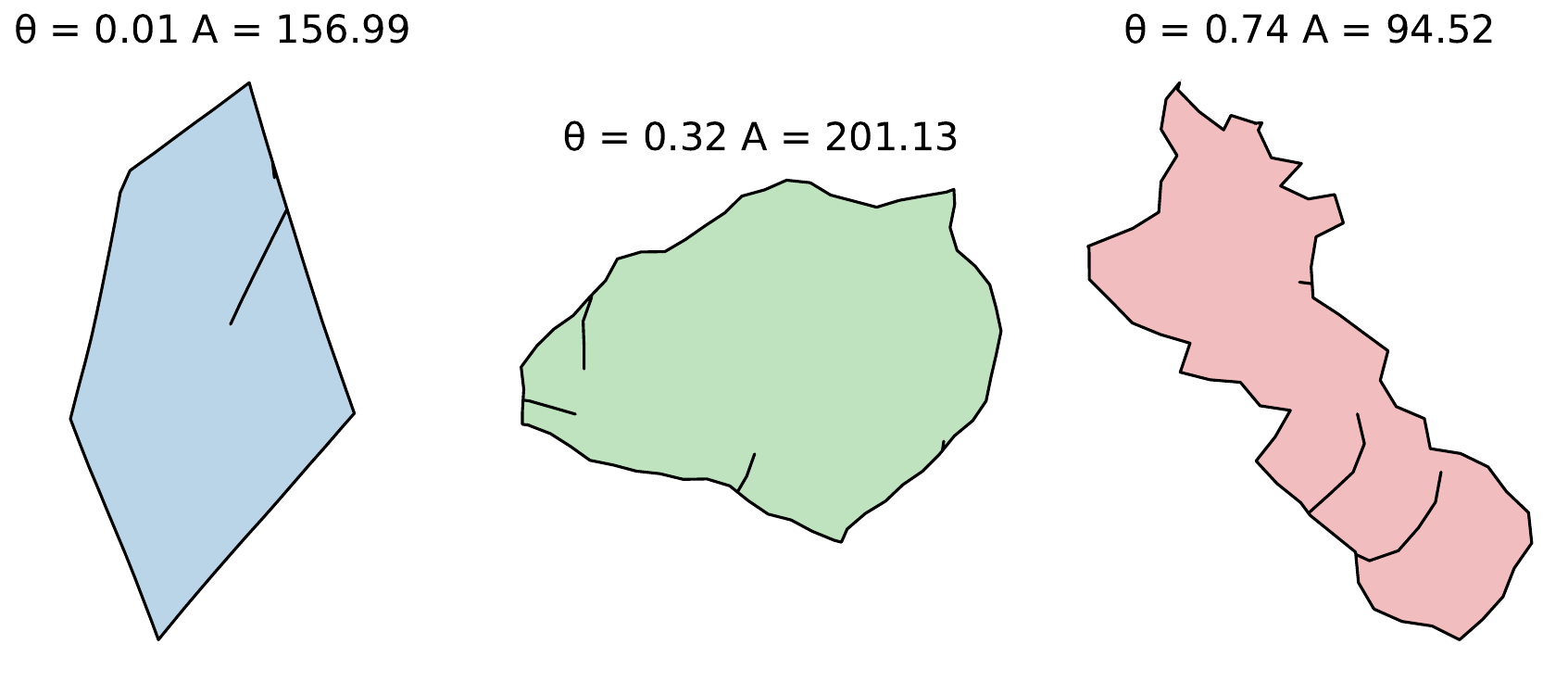}
    \caption{Example of large faces for~$\theta=0.01$,~$\theta=0.32$ and~$\theta=0.74$.}
    \label{fig:largefaces}
\end{figure}

\begin{figure}[H]
    \centering
    \includegraphics[width=1\linewidth]{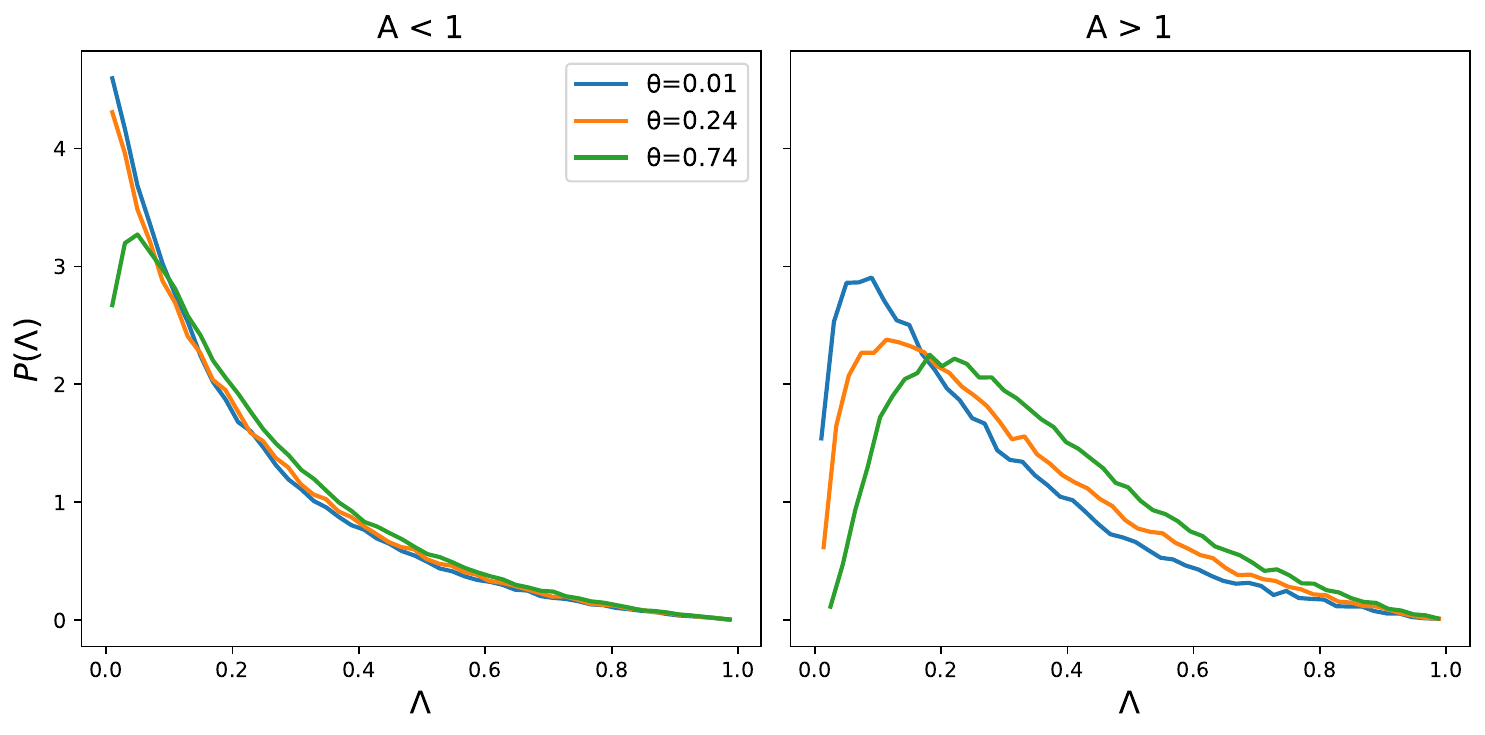}
    \caption{Conditional distribution of anisotropy factor~$\Lambda$ on~$A \gtrless 1$. For small faces, the distribution of anisotropy coincide at large~$\Lambda$, while rods-like chains have more~$\Lambda \approx 0$ faces compared to correlated random walks. Instead, large faces ($A>1$) show a shift of the distribution with~$\theta$, demonstrating that the faces become more and more isotropic as the correlated random walks become correlated on smaller scales.}
    \label{fig:anisotropy}
\end{figure}

Connected planar graphs are characterized by the Euler relation~$f=e-n+2$, linking the number of faces~$f$ to the total number of edges. The maximum number of faces in a planar graph is~$2n-5$~\cite{Barthelemy_2011}. Here, we characterize rather the~\textit{size} and~\textit{shape} of typical faces given by the model in the supercritical ($\mu>1$) phase. We find that, in the open boundary case, when divided by their average~$\tilde{A} \leftarrow A/\langle A \rangle$, the distribution of the inverse areas is
\begin{equation}
    P(1/\tilde{A}) \sim (1/\tilde{A})^{-3/2} \,,
\end{equation}
as shown in Fig.~\ref{fig:area_universal}.
Up to a rescaling factor, the small faces of the model have the same area statistics, independently of~$\theta$. This is not surprising when considering that the length scale at which curvature appears is too large to influence the small faces. Instead, for large faces (illustrated in Fig.~\ref{fig:largefaces}), we find different shapes, characterized by the anisotropy factor~\cite{Kirkley_2018}
\begin{equation}
    \Lambda = \frac{\lambda_{m}}{\lambda_{M}} \,,
\end{equation}
where~$\lambda_m$ ($\lambda_M$) is the smallest (largest) eigenvalue of the covariance matrix of the vertices. $\Lambda$ ranges between~$0$ and~$1$ and quantifies and measure the relative width of the sites characterizing a face. Small values ($\Lambda\sim0$) indicate faces with a one-dimensional layout, while as~$\Lambda$ increases, faces are more and more isotropic.
Fig.~\ref{fig:anisotropy} shows the the distribution of~$\Lambda$, conditioned on~$A<1$ and~$A>1$. As~$\theta$ increases, the shape of large faces become more and more isotropic, as shown by \ulysse{an increase of the typical~$\Lambda$ value}, reflecting the collective effect (as faces appear as junction of many chains) of crossings of smoother chains.

\clearpage

\section{Navigability} \label{sec:navigability}

Aside geometry, networks also induce a discrete topology. Contrarily to standard complex networks, networks with strong spatial constraints typically show strongly bounded degree distributions, and their topological structure is rather assessed through other observables~\cite{Barthelemy_2011}. One of the simplest quantity  is the betweenness centrality
\begin{equation}
    g(v) = \sum_{i \neq j} \frac{\sigma_{ij}(v)}{\sigma_{ij}}
\end{equation} 
\ulysse{
where~$\sigma_{ij}(v)$ counts the number of shortest paths between~$i$ and~$j$ passing through the node~$v$ and~$\sigma_{ij}$ the total number of shortest paths between those nodes.~$g(v)$ thus ranges between~$0$ and~$n(n-1)$.} The BC values of nodes for networks sampled from the chain-network nodel are shown in Fig~\ref{fig:fig:open_boundaries}. It is an essential proxy to topological importance and potential congestion. In large complex networks, it has been shown that BC scales with the node degree as~$g \sim k^{\eta}$ revealing strong structural heterogeneity and the emergence of central hubs~\cite{Barthelemy_2004}. Narayan and Saniee~\cite{Narayan_2011} demonstrated that betweenness centrality reflects an intrinsic large-scale \textit{negative curvature}~\cite{Gromov1987} in complex networks, where the load at the core scales as $n^2$--in contrast to the $n^{3/2}$ scaling expected for flat geometries--thereby revealing how network hyperbolicity~\cite{Boguna2010,krioukov2010} fundamentally amplifies congestion.

In spatial networks, Gago et al.~\cite{Gago2012} demonstrated that the average BC scales proportionally to the mean shortest-path length, underscoring its role as a global efficiency indicator. Kirkley et al.~\cite{Kirkley_2018} found that planar graphs display an invariant BC distribution, explained by the emergence of a bimodal regime, with high BC nodes belonging to the network backbone while additional edges provide alternate pathways, leading to small BC. The BC spatial configuration was firstly investigated in infinite density limit~\cite{Giles_2015}, and then in the large but finite density case~\cite{Verbavatz_2022}. 

\begin{figure}[H]
    \centering
    \includegraphics[width=1\linewidth]{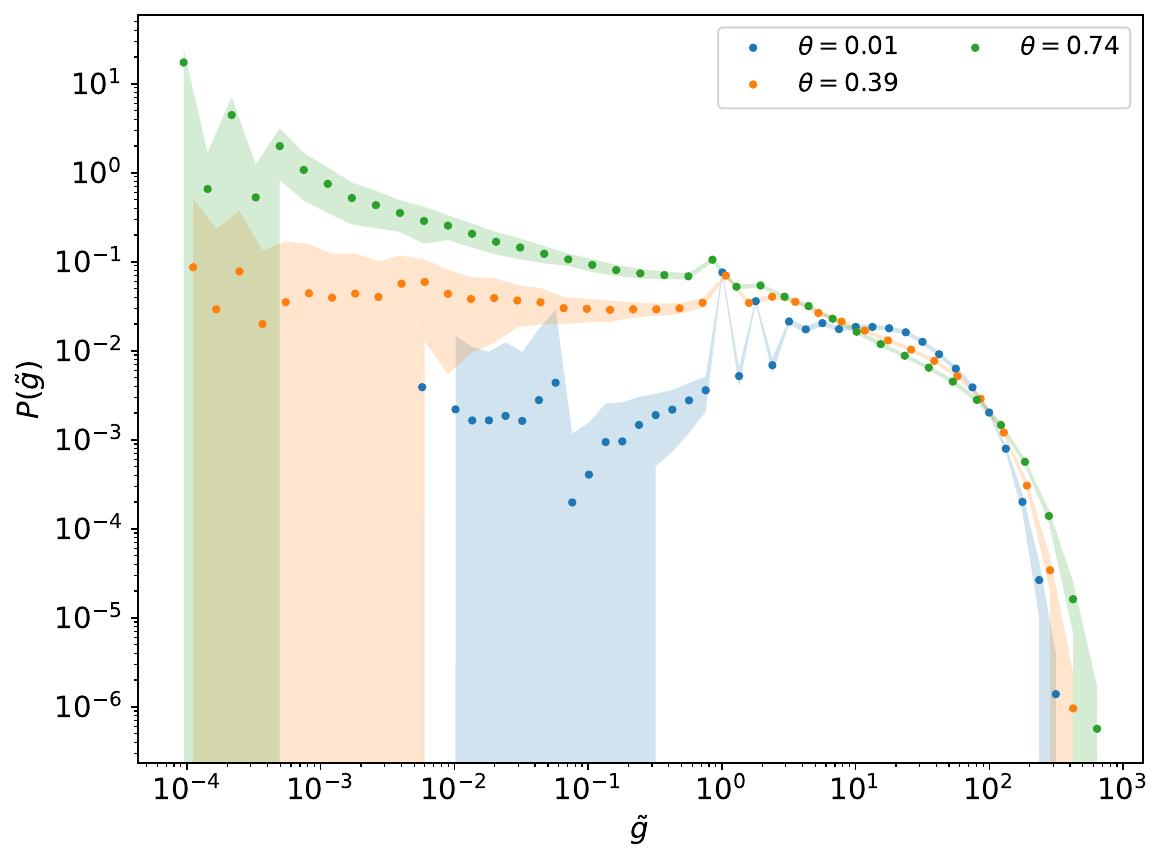}
    \caption{\ulysse{Histogram averaged over~$100$ realizations} (with~$95\%$ error bars) of normalized betweenness centrality for chain-networks \ulysse{with open boundaries}. The expected peak around~$\tilde{g}\approx1$ is present, as well as the predicted bimodal regime.}
    \label{fig:density_bc}
\end{figure}

Given the planarity of networks in the ensemble, the loop density is expected to control the BC distribution. In Fig.~\ref{fig:density_bc} is displayed the BC distribution for different values of~$\theta$. As anticipated, we observe a peak around $\tilde{g}=g/n\approx1$, characteristic of planar graphs~\cite{Kirkley_2018} for all cases. However, all networks have a large number of loops (and a similar amount of loop per node)  but show a completely BC profile, due to the loop localization. The branches-and-center organization of the rods-like systems, for which there are very few peripheral loops, cancels the appearance of~$\tilde{g}<1$ nodes, yield a quasi-unimodal distribution, while for systems made of RW-like chains, the large amount of loops both in the core of the network and in periphery leads to a multiplicity of paths everywhere and hence the BC distribution in this case resembles the familiar bimodal distributions of~\cite{Kirkley_2018}.

\begin{figure*}[h]
    \centering
    \includegraphics[width=1\linewidth]{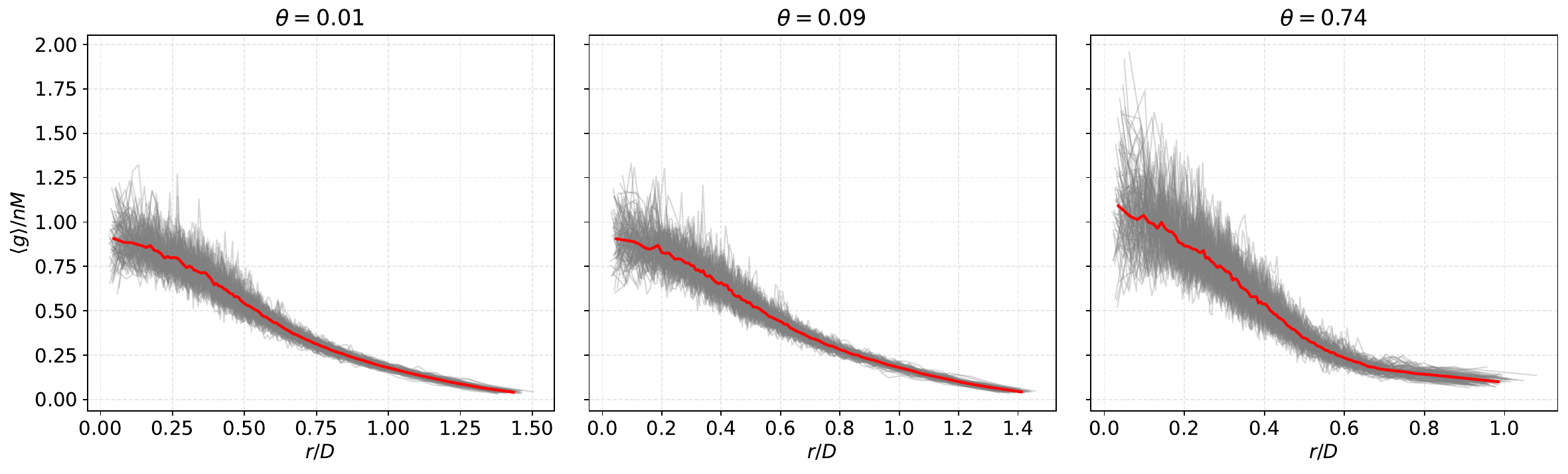}
    \caption{Radial BC distribution, in the open-boundary case. The abscissa is normalized by the domain lateral size~$D$, while the ordinate is divided by ~$nM$. Grey curves represent one realization each, while the red curve is their average.}
    \label{fig:radial_bc}
\end{figure*}

In Fig.~\ref{fig:radial_bc}, we show the radial distribution of betweenness distribution~$g(r)$ computed on chain-network realizations with open boundaries (as boundaries influence strongly BC measurements~\cite{gil2017}) normalized on abscissa by~\ulysse{$\Delta$} and in ordinate by~$nM$, for comparison's sake. 
The resulting profiles further display the importance of peripheral loops on the navigability of the systems : far from the center, the average BC does not vanish in the case of ~$\theta=0.74$ while it does for~$\theta=0$. Another fact is that fluctuations around the average disappear far from center in the case of rods, while they are stronger in the case of curved CRWs. Finally, the average central ($r=0$) normalized BC is higher at the center for curved CRWs, sign of a higher central load on the network and less efficient circulation in those configurations.

\clearpage

\section{Scaling of~$T$ and~$D$ for~$\theta \in [0.1,0.7]$} \label{sec:intermed}
\begin{figure}
    \centering
    \includegraphics[width=0.8\linewidth]{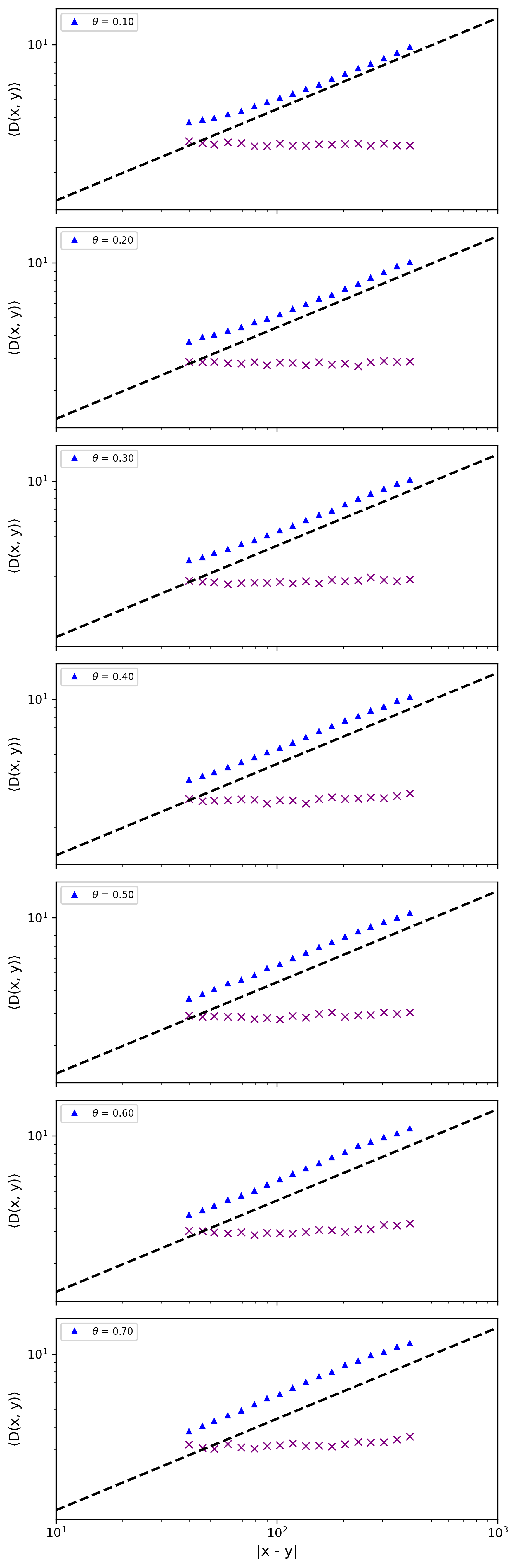}
    \caption{\ulysse{Scaling of~$\mathrm{E}[D]$ (blue triangles) and~$V(T)$ (purple crosses) as a function of the distance, for~$\theta$ ranging between~$\theta=0.1$ and~$\theta=0.7$ (top to bottom). The dark dashed lines represent a power-law of exponent~$1/2$.}}
    \label{fig:scaling_interm}
\end{figure}

\ulysse{
Fig.~\ref{fig:scaling_interm} shows the statistics of the observables~$T$ and~$D$ measured, with the same experimental setup a sdescribed in~\ref{subsec:expe}, for values of~$\theta$ ranging in~$[0.1,0.7]$. The same scaling behaviors as in the cases~$\theta=0, 0.09, 0.74$ are retrieved. Note the reduction of the initial crossover regime as ~$\theta$ increases. The exponents are estimated using a linear fit in log-log scale, after excluding the crossover regime from the scaling range. The measured exponents lie in the range~$[0.47,0.5]$ for~$\xi$ and in the range~$[-0.01,0.04]$ for~$\chi$, coinciding with the results presented in~\ref{subsec:expo}.
}

\clearpage

\section{Variance of the transversal deviations}

\begin{figure}[H]
    \centering
    \includegraphics[width=0.9\linewidth]{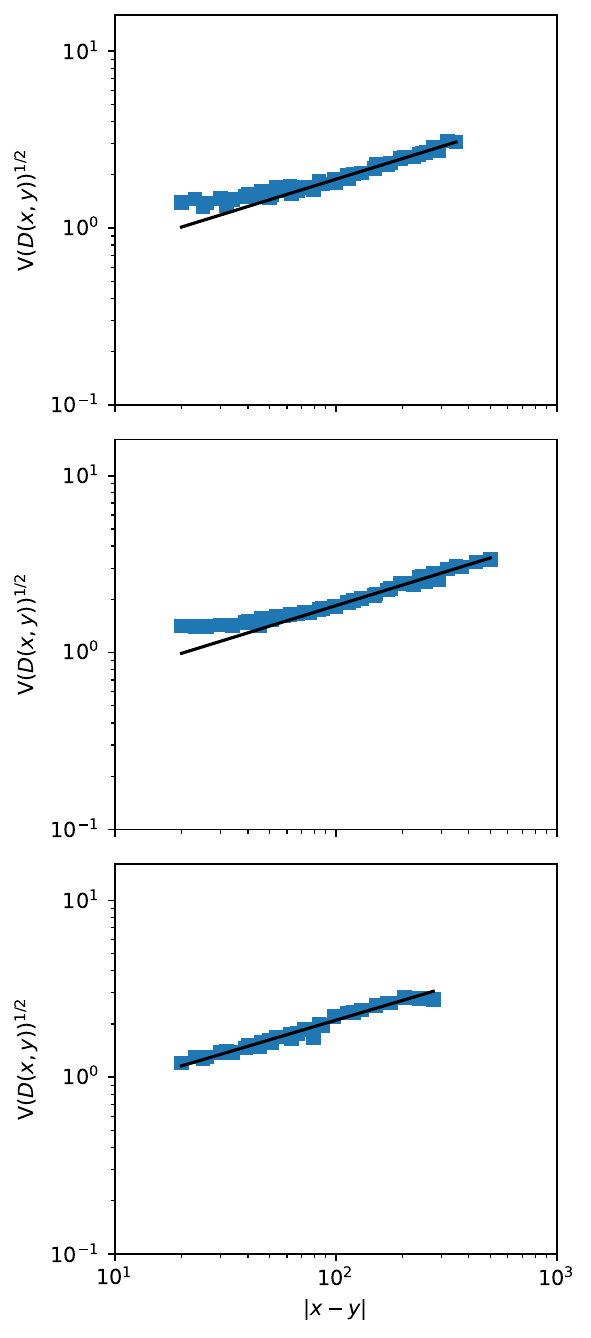}
    \caption{Scaling of standard deviation of the transverse deviations with Euclidean distance for~$\theta=0$ (top),~$\theta=0.09$ (middle) and~$\theta=0.74$ (bottom). Black lines : power-law fits of exponent~$0.39$, $0.39$ and $0.37$ (up to down).}
    \label{fig:std_transverse}
\end{figure}

Fig.~\ref{fig:std_transverse} shows the standard deviation of~$D$,~$V(D)^{1/2}$, in function of the Euclidean distance.

\clearpage


%

\end{document}